\documentclass[structabstract]{aa}

\usepackage{adjustbox}
\usepackage{color}
\usepackage{graphicx}
\usepackage{amsmath}
\usepackage[table]{xcolor}

\usepackage{natbib}
\usepackage{graphicx}
\usepackage{rotating}
\usepackage{txfonts}
\usepackage{lscape}
\usepackage{booktabs}
\usepackage{dcolumn}
\usepackage{float}
\usepackage{dblfloatfix}
\usepackage{array}
\usepackage{footnote}
\usepackage{wasysym}
\makesavenoteenv{tabular}
\makesavenoteenv{table}
\usepackage{diagbox}

\usepackage{txfonts}
\usepackage[colorlinks=true,
            linkcolor=blue,
            citecolor=magenta]{hyperref}
            
\title{Probing radiation micro-physics in M\,87 \\ I. Total intensity and broad-band spectra}\titlerunning{Probing radiation micro-physics in M\,87}

\author{Christian M. Fromm \inst{\ref{affil:wuerzburg},\ref{affil:frankfurt}}\fnmsep\thanks{christian.fromm@uni-wuerzburg.de}
\and Yosuke Mizuno \inst{\ref{affil:tsung},\ref{affil:jiaotong},\ref{affil:keylab},\ref{affil:frankfurt}}
\and Ziri Younsi \inst{\ref{affil:mullard}}
\and Ainara Saiz-P\'{e}rez \inst{\ref{affil:wuerzburg}}
\and Hector Olivares \inst{\ref{affil:aveiro}}
\and Antonios Nathanail \inst{\ref{affil:athens}}
\and Alejandro Cruz-Osorio \inst{\ref{affil:unam}}
\and Matthias Kadler \inst{\ref{affil:wuerzburg}}
\and Karl Mannheim \inst{\ref{affil:wuerzburg}}}
\institute{%
  Institut f\"ur Theoretische Physik und Astrophysik, Universit\"at W\"urzburg, Emil-Fischer-Str. 31, D-97074 W\"urzburg, Germany\label{affil:wuerzburg}
  \and%
  Institut f\"ur Theoretische Physik, Goethe Universit\"at, Max-von-Laue-Str. 1, D-60438 Frankfurt, Germany \label{affil:frankfurt}
  \and 
  Tsung-Dao Lee Institute, Shanghai Jiao Tong University, 1 Lisuo Road, Shanghai 201210, China \label{affil:tsung}
  \and
  School of Physics \& Astronomy, Shanghai Jiao Tong University, 800 Dongchuan Road, Shanghai 200240, China \label{affil:jiaotong}
  \and
  Key Laboratory for Particle Physics, Astrophysics and Cosmology, Shanghai Key Laboratory for Particle Physics and Cosmology, Shanghai Jiao-Tong University, 800 Dongchuan Road, Shanghai 200240, China\label{affil:keylab}
 \and 
 Mullard Space Science Laboratory, University College London, Holmbury St. Mary, Dorking, Surrey RH5 6NT, UK \label{affil:mullard}
 \and Instituto de Astronom\'{\i}a, Universidad Nacional Aut\'onoma de M\'exico, AP 70-264, Ciudad de M\'exico 04510, M\'exico \label{affil:unam}
 \and Research Center for Astronomy and Applied Mathematics, Academy of Athens, GR 11527 Athens, Greece \label{affil:athens}
 \and Departamento de Matematica da Universidade de Aveiro and Centre for Research and Development
in Mathematics and Applications (CIDMA), Campus de Santiago, 3810-193 Aveiro, Portugal \label{affil:aveiro}
  }%

\begin{document} 

\abstract
  {Next generation Very Long Baseline Interferometers (VLBI) will provide dense sampling of the 
  Fourier space together with high signal to noise ratios allowing to {reliably} observe and image 
  faint jet structure in M\,87 at mm-wavelength. The proposed next generation Event Horizon 
  Telescope (ngEHT) and next generation Very Large Array (ngVLA) offers the unique capability to simultaneously resolve and image the 
  accretion flow around the supermassive black hole in M\,87 together with the jet launching 
  and acceleration zone.  In order to explore these capabilities and to provide theoretical expectations 
  we perform general relativistic magnetohydrodynamic simulations of accretion on to black 
  holes and jet launching.}
  {M\,87 has been the target for multiple observations across the entire electromagnetic spectrum. 
  Among these VLBI observations provide unique capability to resolve the jet structure  down to 
several gravitational radii. In this work we provide possible observable signatures which will allow us to
 distinguish between different electron heating models and particle distributions.}
  {We use general relativistic magnetohydrodynamics and simulate the accretion of the magnetised
   plasma onto Kerr-black holes in 3D. The multi-frequency radiative signatures of these simulations are computed 
taking different electron heating and distribution functions into account.}
  {The results of our simulations show that with a dynamical range of $1\times 10^4$ and a frequency range from 86\,GHz to 345\,GHz observations with future VLBI arrays have the potential to tell turbulent and magnetic reconnection electron heating and the electron distribution function apart.}
  {}

\maketitle

\section{Introduction}\label{sec:intro}
The elliptical galaxy M\,87 with its high-mass supermassive black {hole} of $M_{\rm BH}=6.5^{+0.7}_{-0.7}
\times 10^{9} M_{\odot}$  and its close proximity of $\sim$16.9 Mpc \citep{EHT_M87_PaperI} is a marvellous object to study  
accretion onto black holes, the formation of relativistic jets and the acceleration of particles. The horizon structure of M87 has been observed at 230\,GHz by the Event Horizon Telescope (EHT) in total intensity \citep{EHT_M87_PaperI, EHT2024_M87} and in polarised light \citep{EHT2021_M87pol}. The analysis of these observations revealed an asymmetric ring of
diameter $d=\, 43\,^{+1.5}_{-3.1}\, {\mu \rm as}$ and a coherent azimuthal polarisation structure with polarisation fraction up to 15\%.
In addition to the EHT observations \citep{Lu2023} observed M\,87 at 86\,GHz using the Global Millimetre VLBI array (GMVA) including the ALMA and the Greenland telescope. These observations showed for the first time a ring like structure together with an {extended} jet. However the extracted ring at 86\,GHz is roughly 50\% larger than the ring at 230\,GHz. The reconstructed jet at 86\,GHz shows strong edge brightened structure together with a broad jet base. These observations {confirmed a} large opening angle of $\phi=63.6^\circ\pm25^\circ $ extracted from 86\, GHz GMVA of \citep{Kim2018a}.  {At frequencies below 86\,GHz, M\,87 has been the target of numerous observing campaigns and long-term monitoring programs.} These observations allowed to extract the jet shape \citep{Hada2013}, magnetic field via the core-shift \citep{Asada2012},  kinematics \citep{Mertens2016,Walker2018} and periodic oscillation of jet base \citep{Cui2023}. The spectral indices between 22\,GHz and 43\,GHz has been studied by \citet{Ro2023}. The authors find a flat spectral index in the core region and fast steepening spectral indices in the jet $\alpha<-2.5$\footnote{assuming $S\propto\nu^\alpha$}. Besides the above mentioned radio observations using the technique of Very Long Baseline Interferometry (VLBI) which allow to resolve and image the horizon and jet structure, M\,87 is frequently observed in the high-energy regime including x- and $\gamma$-rays. Simultaneous multi-wavelength observations in combination with imaging VLBI observations such as \citet{Algaba2021,EHTMW2024} allow to probe the underlying particle acceleration and energy extraction mechanism in M\,87 in particular or in relativistic jets in general. 
\newline Despite substantial progress in the theoretical understanding of accreting black holes and jet launching the detailed mechanism is still on debate. The most promising ones are the energy extraction from a rotating accretion disk or by directly tapping the energy of rotating black holes \cite{Blandford:1982di,Blandford1977,Blandford2019}. The repeated EHT observations of the horizon scale structure of M\,87 and its accompanied modelling via General Relativistic Magneto-HydroDynamic (GRMHD) simulations are in agreement with estimates based on the Blandford-Znajek process \citep{Blandford1977,EHT_M87_PaperV}.  
To draw a self-consistent picture of accretion onto a black hole,
GRMHD simulations together 
with general relativistic radiative transfer (GRRT) calculations
\citep[see, e.g.][]{Dexter2012} are employed. The characteristics of the M\,87
radio core, of a flat spectrum and an
increasing size with wavelength can be reproduced by a two-temperature
accretion flow and a hot single-temperature jet \citep{Moscibrodzka2016}.
The addition of non-thermal electrons is expected to better reproduce the
NIR flux and the flat radio spectrum, but also acquire a more extended jet 
structure at 43\,GHz and 86\,GHz \citep{Davelaar2019}.
\newline In this work we study the impact {of} electron heating models as well as 
of different electron distribution functions (eDF) on the
structure and morphology of the M\,87 jet and its spectrum. To this scope we perform
long-term, high-resolution, 3D GRMHD simulations of magnetically arrested disks  (MAD)
with electron heating around fast spinning Kerr black holes ($a_{\star}=cJ/GM^2=0.9375$). 
Here, we focus only on MAD models given that they better describe/fit the EHT observations \citep{EHT_M87_PaperV, EHT_M87_Paper8}
{While several mechanisms may in principle contribute to plasma heating (e.g. shocks or other dissipation processes), in this work we restrict ourselves to turbulent \citep[see, e.g.,][]{Kawazura2019} and reconnection-based  \citep[see, e.g.,][]{Rowan2017}  heating prescriptions. This choice follows \cite{Mizuno2021} and reflects the current state-of-the-art sub-grid models that can be consistently implemented in GRMHD simulations which allow a controlled and physically motivated comparison. In addition, particle acceleration at shocks is expected to be inefficient in magnetised plasmas, particularly in relativistic, quasi-perpendicular configurations \citep[see, e.g.,][]{Sironi2009}, which further motivates focusing on turbulence and reconnection considered here.} The output of these simulations is coupled with GRRT calculations, 
where we employ a mixture of thermal and non-thermal electrons using the kappa
distribution \citep{Davelaar2019,Osorio2021,Fromm2021}. {The application of the kappa‑distributions to model the heated particles are motivated by kinetic theory and observations, as stochastic acceleration in turbulent/reconnection environments naturally yields kappa‑type equilibria \cite[see, e.g.,][]{Bian2014, Lazar2021}.}

In this work we expand the results presented in 
\citet{Zhang2024} while focusing multi-frequency and spectral properties of high spinning $a_{\star}=0.9375$ two-temperature MAD models.
\newline The paper is organised as follows: In Section \ref{sec:GRMHD} we discuss
the details of the GRMHD simulations, whereas in Section \ref{sec:GRRT}
we describe the GRRT calculations. In Section
\ref{sec:results} we present our results and provide our discussion in Section \ref{sec:discussion}. Lastly, we present a summary and an outlook in Section \ref{sec:conc}.
\newline Throughout this work we use a black hole mass of 
$6.5\times10^9\,M_{\astrosun}$ for M\,87 a distance of 16.8\,Mpc and a viewing angle of $\vartheta=160^\circ$. 

\section{General Relativistic Magneto-Hydrodynamic simulations}\label{sec:GRMHD}
For our numerical modelling of black hole accretion and jet launching we employ the
 state-of-the-art GRMHD code \textit{Black Hole Accretion Code} \texttt{BHAC}
 \cite{Porth2019, Olivares2020}. We initialise our 3D MAD GRMHD simulations with
a magnetised torus in hydrostatic equilibrium where the inner edge of the torus, 
$r_{\rm in}$, is located at $20\,M$ and location of the pressure maximum can be found
at $r_c=40\,M$ {and the outer boundary of the torus is located at $r \sim 400\,M$, beyond which a floor model is applied (see below).} The initial torus is filled with a single looped poloidal magnetic field 
given by the vector potential:
\begin{equation}
   A_{\phi}\propto\mathrm{max}(q-0.2,0)\, {\rm with}\, q = \frac{\rho} {\rho_{\rm max}}\left(\frac{r}{r_{\rm in}}\right)^3 \sin^3 \theta \exp \left(\frac{-r}{400}\right)
\end{equation} 
see, e.g., \cite{Fishbone76,Font02b,Rezzolla_book:2013}.

In addition to the standard GRMHD quantities such as density, $\rho$, and four magnetic field, $b^\mu$, we additionally evolve the electron entropy where the heating fraction of the electrons, $f_e$, for turbulent heating and magnetic reconnection heating are included. {The heating fractions can be written as:
\begin{eqnarray}
f_e &=& \frac{1 + (\beta_{\rm p}/15)^{-1.4}\,\exp\!\left(-0.1\,\frac{T_e}{T_i}\right)}{1 + (\beta_{\rm p}/15)^{-1.4}\,\exp\!\left(-0.1\,\frac{T_e}{T_i}\right) + 35} \quad \rm{turb.\,heating}
\label{eq:turbheating} \\
f_e &=& \frac{1}{2}\,\exp\!\left[-\,\frac{1 - \beta_{\rm p}/\beta_{p,\max}}{0.8 + \sigma_h^{1/2}}\right] \quad \rm{reconnection\, heating}
\label{eq:reconheating}
\end{eqnarray}
where $\beta_{p,\max} = {\sigma_h}/{4}$, $\sigma_h = {b^2}/{(\rho h)}$ is the magnetisation with respect to the specific enthalpy $h = 1 + {\hat\gamma}p/({\hat\gamma- 1})$. The turbulent heating prescription of \cite{Kawazura2019} assumes a quasi-steady Alfvénic cascade and neglects compressive fluctuations, which are expected to predominantly heat ions, especially at low plasma beta. In addition, the prescription depends primarily on $\beta$ (and the temperature ratio) and does not explicitly account for the magnetisation ($\sigma$), which may become important in strongly magnetised regions. The reconnection-based model of \cite{Rowan2017} is calibrated on PIC simulations of idealised current sheets (e.g. anti-parallel configurations with controlled plasma parameters) and thus does not capture the full range of geometries, variability, or non-local effects in global flows. 
Despite these simplifications, both prescriptions capture the dominant dependencies of electron heating on local plasma conditions and provide a physically well-motivated framework for incorporating kinetic effects in GRMHD simulations.For more details on the numerical implementation and convergence of the heating models see \cite{Mizuno2021}.} Throughout this work we use an adiabatic index of $\hat{\gamma}=4/3$.  

{As is standard in GRMHD simulations, vacuum regions are avoided by introducing a low-density ``atmosphere'' \citep{Rezzolla_book:2013}. In practice, floor values are imposed on the rest-mass density and gas pressure, $\rho_{\rm fl} = 10^{-4} r^{-2}$ and $p_{\rm fl} = (10^{-6}/3)\, r^{-2}$, such that in cells where $\rho \le \rho_{\rm fl}$ or $p \le p_{\rm fl}$, the primitive variables are reset to these floor values. This treatment is required for numerical stability and is particularly relevant in low-density regions such as the jet spine, where the physical density can become extremely small. As a result, these regions are typically characterised by high magnetisation ($\sigma > 1$) and should be interpreted with care, as the imposed floors can affect thermodynamic quantities, including the inferred electron temperature and heating properties. For subsequent analysis and GRRT post-processing, such regions are therefore commonly excluded by applying a $\sigma$ cut-off \citep[see, e.g.,][]{EHT_M87_PaperV,Fromm2021}. Furthermore, we impose both a floor and a ceiling on the electron pressure to maintain numerical stability. In practice, if $p_e < 0.01\,p_{\rm fl}$, we set $p_e = 0.01\,p_{\rm fl}$, while if $p_e > p_g$, we set $p_e = 0.99\,p_g$, which in turn imposes corresponding bounds on the electron temperature \citep[for details see][]{Mizuno2021}}

{Notice, that the initial condition used in this work are motivated by previous modelling of EHT observations of M87 \cite{EHT_M87_PaperV}. While this setup provides a well-established and observationally grounded framework, we note that the resulting turbulent and reconnection properties and thus the detailed heating may depend on the choice of initial conditions (e.g., torus size, magnetic field configuration, or MAD vs. SANE state, adiabatic index). Exploring this dependence remains an important direction for future work beyond the scope of this study.}

\begin{figure}[h!]
\centering
\includegraphics[width=0.49\textwidth]{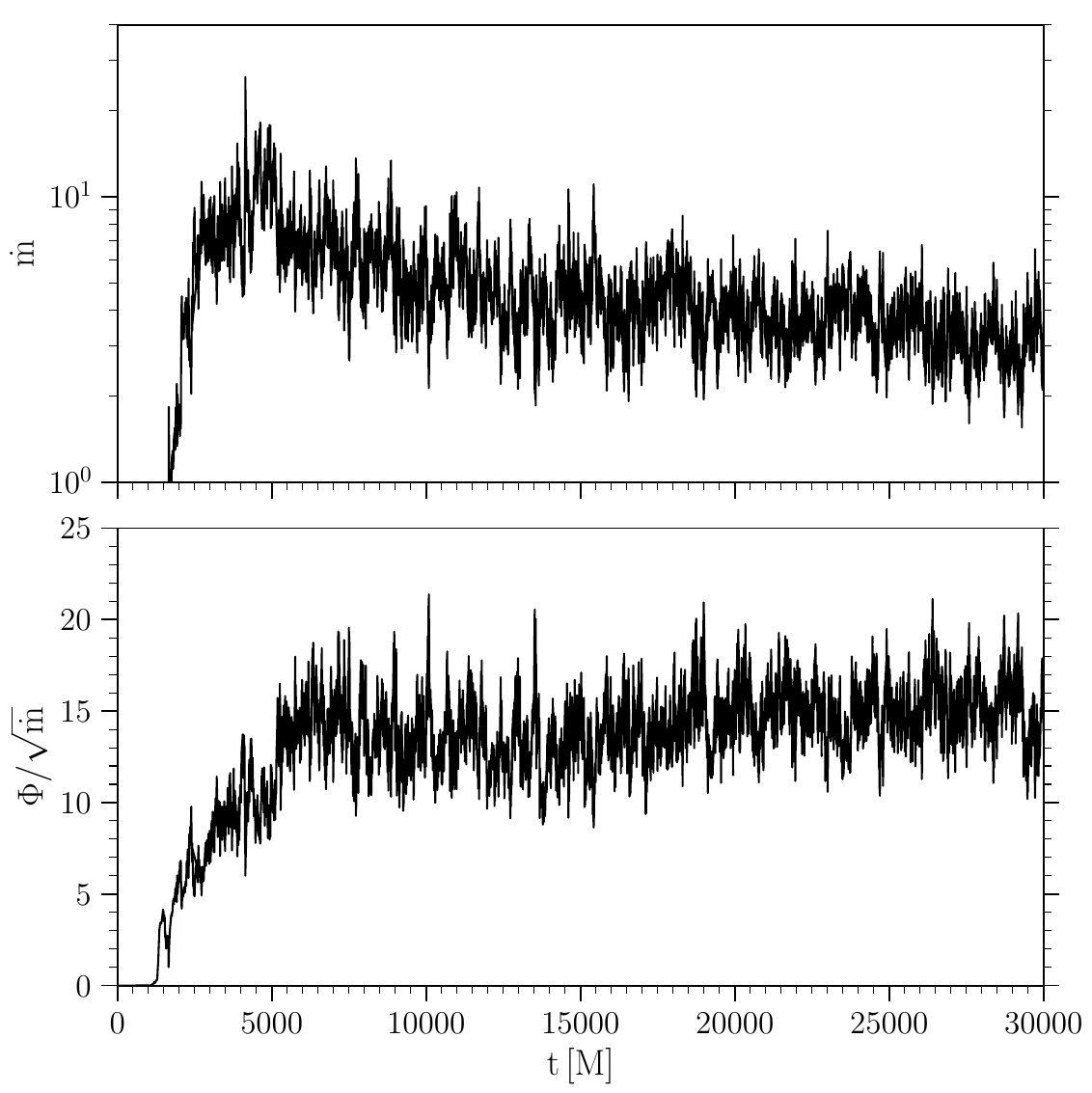}
 \caption{Mass accretion rates $\dot{m}$ and MAD flux parameter $\phi=\Phi/\sqrt{\dot{m}}$ in code units, where $\Phi$ is the magnetic flux across the horizon, for a MAD GRMHD simulation with $a_{\star}=0.9375$.}
\label{fig:rates}
\end{figure}

Notice, that for each heating model a individual GRMHD simulations is performed. The simulations are performed on spherical grid $(r,\theta, \phi)$ using logarithmic Kerr-Schild coordinates and employing three layers of adaptive mesh refinement (AMR) leading to an effective resolution of $384\times192\times192$. The outer radial boundary of the numerical grid is located at 2500\,M. To start the accretion onto the black hole we trigger the magneto-rotational instability (MRI) by perturbing the pressure of equilibrium torus with white noise $p = p (1 + X_p)$ where $|X_p|<0.04$ is a random number. During the course of the GRMHD simulations we monitor the mass accretion rate, $\dot{m}$, and the accreted magnetic flux across the horizon, $\Phi$. In Figure \ref{fig:rates} we present the evolution of the mass accretion rate, $\dot{m}$, and the MAD flux parameter, $\phi=\Phi/\sqrt{\dot{m}}$, for the MAD model
with $a_{\star}=0.9375$. Since the GRMHD parameters are not altered by the change in the electron heating only one set of values for $\dot{m}$ and $\Psi$ are presented\footnote{Variation in the accretion parameters are due to different turbulent realisations within the plasma.}. After $t\sim 10,000\,\rm M$ a  {quasi-steady} mass accretion rate is obtained and within the time interval of  $10000\,\rm{M} \leq t \leq 30000\,\rm{M}$ we obtain an average mass accretion rate of $\langle \dot{m}\rangle=3.98\pm1.15$ and an average MAD parameter of $\langle \phi\rangle=14.4\pm1.85$. With $\langle\phi\rangle\sim14$ our  model is well within the MAD state according to the threshold value of $\phi \sim 10$ \citep{Tchekhovskoy2011}. 

\begin{figure}[h!]
\centering
\includegraphics[width=0.49\textwidth]{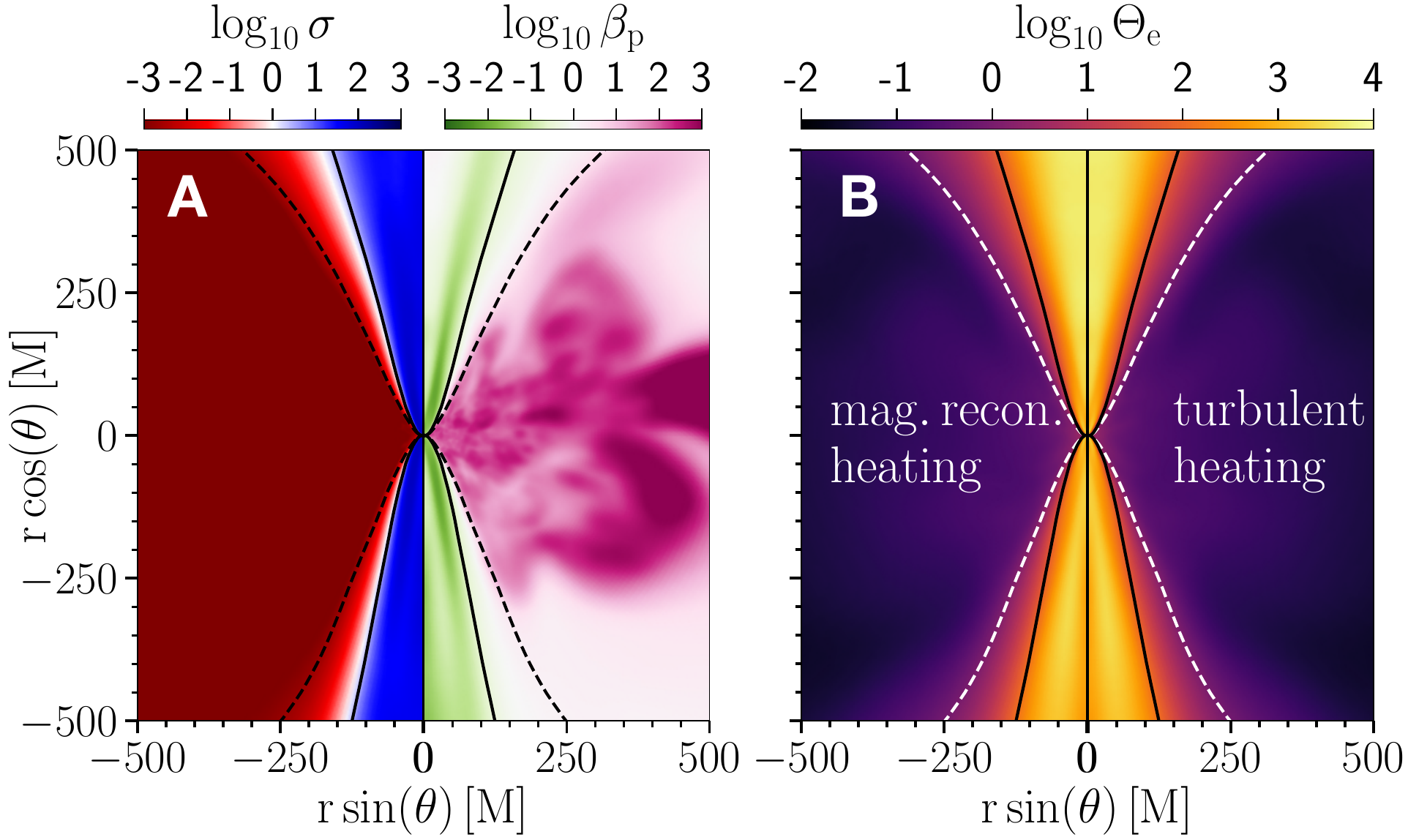}
 \caption{{Azimuthally and time-averaged ($28000\,\mathrm{M} \leq t \leq 30000\,\mathrm{M}$) distributions of the magnetisation $\sigma$ (left half of panel A), plasma beta $\beta_{\rm p}$ (right half of panel A), and electron temperature for magnetic reconnection heating (left half of panel B) and turbulent heating (right half of panel B). Solid black lines mark $\sigma = 3$, while dashed lines indicate the boundary between bound ($-h u_t < 1.02$) and unbound ($-h u_t > 1.02$) plasma.}}
\label{fig:Morphology1}
\end{figure}

{The azimuthal and time averaged (using the interval $28000\,\rm{M} \leq t \leq 30000\,\rm{M}$) distribution of the magnetisation of the plasma, $\sigma$, plasma-beta, $\beta_{\rm p}$ and the dimensionless electron temperature, $\Theta_{\rm e}$, are presented in panel A and B in Fig. \ref{fig:Morphology1}. We show the azimuthal and temporal averaged distribution of $\sigma$ in left half of panel A in Fig.\ref{fig:Morphology1} and for $\beta_{\rm p}$ in the right half of panel A in Fig.\ref{fig:Morphology1}. As expected for MAD models our simulations exhibit a large opening angle, indicated by the dashed black line which trace the Bernoulli parameter ($-hu_{t}=1.02$), and show a highly magnetised low plasma-beta jet region. At the same time the disk region ($-hu_{t}<1.02$) is characterised by large plasma beta values, $\beta_{\rm p}>0.1$, and low magnetisation, $\sigma<0.01$. The solid black line indicates the $\sigma=3$ contour, i.e. the region affected by the floor model.}

\begin{figure*}[b!]
\centering
\includegraphics[width=\textwidth]{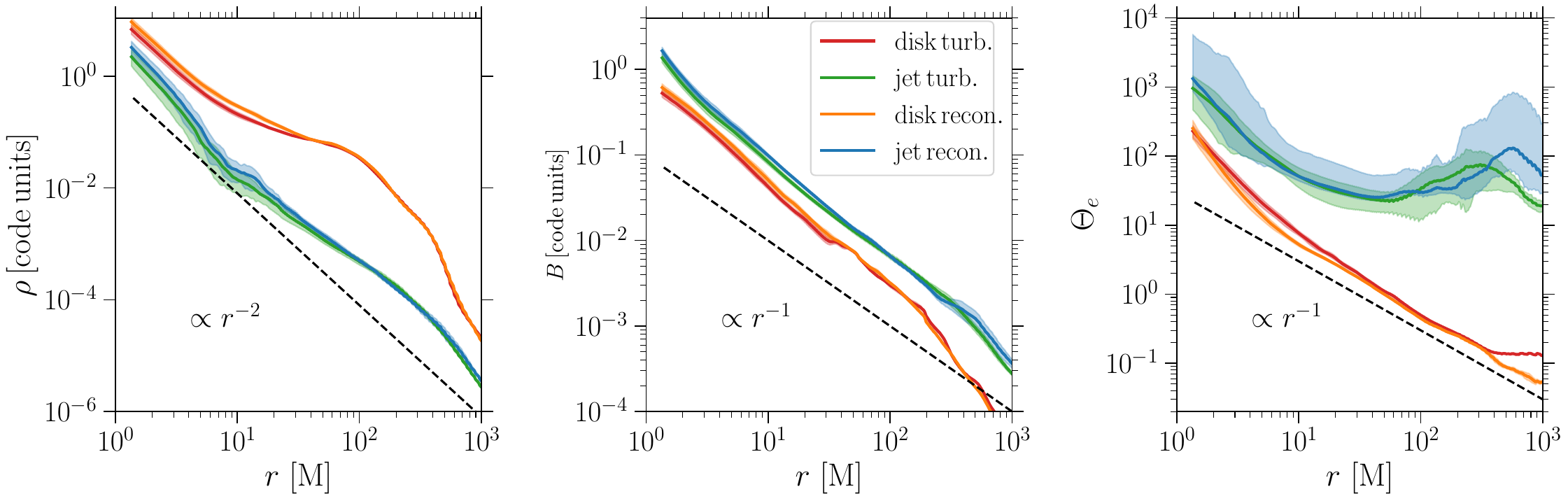}
 \caption{{Time-averaged radial profiles over $28000\,\mathrm{M} \leq t \leq 30000\,\mathrm{M}$ showing, from left to right, the rest-mass density $\rho$, magnetic field strength $B$, and dimensionless electron temperature $\Theta_{\rm e}$ for both turbulent and magnetic reconnection heating. Profiles are further decomposed into contributions from the disk and jet, while the shaded bands indicate the corresponding $1\sigma$ variability.}}
\label{fig:profiles}
\end{figure*}

{However, due to the two different electron heating models (see Eqn. \ref{eq:turbheating} and \ref{eq:reconheating}) the distribution of the dimensionless electron temperature, $\Theta_{\rm e}$ is different across our models. In panel B of Fig.~\ref{fig:Morphology1} we show in the left half the electron temperature using magnetic reconnection heating and in the right half the electron temperature due to turbulent heating. In the turbulent model, for a fixed temperature ratio $T_{\rm e}/T_{\rm i}$, the electron heating fraction has a strong inverse dependence on $\beta_{\rm p}$, i.e. $f_e \to 1$ for $\beta_{\rm p} \ll 1$ and $f_e \ll 1$ for $\beta_{\rm p} \gg 1$. This leads to efficient electron heating in the low-$\beta_{\rm p}$ jet/wind and ion-dominated heating in the high-$\beta_{\rm p}$ disk. As a result, the turbulent prescription produces a strongly stratified $\Theta_e$ distribution following closely the distribution of $\beta_{\rm p}$ (compare to right half of panel A), with high electron temperatures confined to the funnel and sheath, while the disk mid-plane remains comparatively cold in electrons. \newline  For the reconnection prescription, the electron heating fraction depends primarily on the magnetisation $\sigma$. In the jet and funnel regions ($\sigma \gg 1$), one obtains $f_e \sim 0.5$, leading to near equipartition between electron and ion heating and thus efficient electron heating. In the disk mid-plane ($\sigma \ll 1$, $\beta \gg 1$), the electron heating fraction is significantly reduced ($f_e \ll 1$), and the dissipated energy is predominantly transferred to ions. Compared to the turbulent case, this produces a somewhat broader and more diffuse high-temperature structure within the low-$\beta$ funnel and jet sheath. However, the heating remains largely confined to magnetically dominated regions, and does not extend significantly into the high-$\beta$ disk, where $f_e \ll 1$ and ion heating dominates. \newline The largest differences between the models occur in the jet spine (highest-$\sigma$ regions between the  black solid curves in panel B), where the reconnection prescription shows higher electron temperatures. These regions are, however, affected by numerical floor treatments and are typically excluded in the GRRT post-processing through a $\sigma$ cut-off. As a result, while the intrinsic differences are most pronounced there, their direct observational impact is reduced, and the comparison is primarily driven by the surrounding jet and disk structure. Notice, that  North-South asymmetries in the distributions persist even after time and azimuthal averaging. These asymmetries arise from the stochastic nature of turbulence, inherently asymmetric development of large-scale structures (e.g., magnetic flux bundles and current sheets)  during MAD states and finite averaging intervals. These asymmetries are further enhanced in the low-density jet spine where floor prescriptions can amplify residual fluctuations.}

In order to inspect the radial dependence of {density}, the magnetic field and electron temperature we computed the following integral:
\begin{equation}
\left < q(r) \right >=\frac{1}{\Delta t} \int_{t_{\rm min}}^{t_{\rm max}}\frac{\int_{-\pi}^{\pi}\int_0^{2\pi}q(t,r,\theta,\phi)\sqrt{-g(r,\theta)}d\phi d\theta}{\int_{-\pi}^{\pi}\int_{0}^{2\pi}\sqrt{-g}d\phi d\theta}dt,
\end{equation}
where $q(t,r,\theta,\phi)$ is the quantity to average, and $\sqrt{-g(r,\theta)}$ is the determinant of the metric. The results for the {density}, magnetic field and the electron temperature, $\Theta_{\rm e}$ {using $t_{\rm min}=28000\,M$ and $t_{\rm max}=30000\,M$} are presented in Fig. \ref{fig:profiles}. In addition we separated simulation into disk and jet contribution. We define as disk the region where $-hu_t <1.02$ and magnetisation $\sigma<0.1$ while for the jet we use $-hu_t >1.02$ and $0.1\leq\sigma\leq3.0$ (in agreement with the definition used in Fig. \ref{fig:Morphology1}). The high magnetised outflowing region, e.g., $-hu_t >1.02$ and $\sigma>3$ we define as jet spine. The density in the jet decreases as $r^{-2}$ in agreement with analytical works \citep[see, e.g.,][]{Blandford1979} whereas the disk exhibits is similar behaviour with a plateau between $40\,M\leq r  \leq 100\,M$. The magnetic field for both disk and jet decays like $r^{-1}$ which indicates the dominance of toroidal field in both components in agreement with the results of \citet{Davelaar2019,Porth2019}. However, in the electron temperature both components can be clearly distinguished.  { The electron temperature in the disk, $\Theta_{\rm e}$, shows an approximate $r^{-1}$ dependence and only small variations between the heating models. Since in the disk both prescriptions are applied to high-$\beta_{\rm p}$, low-$\sigma$ regime they lead to $f_e \ll 1$. In this limit, electron heating is strongly suppressed and most of the dissipated energy is transferred to ions. Thus, $\Theta_{\rm e}$ is largely insensitive to the differing $\beta_{\rm p}$ and $\sigma$-dependencies of the used heating prescriptions.}The jet follows the same radial decrease up to 10\,M. For larger distance the jet shows nearly a constant electron temperature of $\Theta_{\rm}\sim 40$ with a slight increase with distance for $r>200\,M$. The trends in the jet are similar for both heating mechanism where the magnetic reconnection heating exhibits a larger variation in the electron temperature as the turbulent heating model. {Note the difference in electron temperature within the disk at large radii ($r > 500\,M$), which lie beyond the extent of the initial torus ($r \sim 400\,M$). Outside the initial torus ($r \gtrsim 400\,M$), the magnetic field is negligible, such that $\beta \to \infty$ and $\sigma \to 0$, and both heating prescriptions predict inefficient electron heating ($f_e \ll 1$). In the turbulent case, the stronger $\beta$-dependence leads to a rapid decline in $\Theta_e$, such that the imposed electron pressure floor produces the observed flattening, while its impact is less pronounced for reconnection heating.}

\section{General Relativistic Radiative Transfer (GRRT) calculations}\label{sec:GRRT}
{We compute the radiative signatures of the GRMHD models using the GRRT code \texttt{BHOSS} \cite{Younsi2020}. In \texttt{BHOSS}, null geodesics are integrated with an RKF45 scheme (fourth-order adaptive stepping, fifth-order error control), while the radiative transfer equations are solved using an Eulerian method with steps provided by the geodesic integration. We adopt a field of view of {$-400,M$ to $400,M$} (corresponding to $\pm1500,\mu\mathrm{as}$ for M87), and a resolution of $8192 \times 8192$ pixels (i.e. $\sim 3$ pixels per $\mu\mathrm{as}$) to simultaneously resolve horizon scales ($r < 100,\mu\mathrm{as}$) and extended jet emission ($r \sim 1000,\mu\mathrm{as}$).
In addition to the two electron heating models, we investigate both thermal and $\kappa$ electron distribution functions (eDFs) and their impact on broadband spectra ($10^{10},\mathrm{Hz} \leq \nu \leq 10^{16},\mathrm{Hz}$), image structure, and spectral properties. We assume synchrotron emission as the dominant process from radio to NIR frequencies and neglect Bremsstrahlung and inverse Compton processes, which only become important above $\nu \gtrsim 10^{16},\mathrm{Hz}$ \citep[e.g.][]{Yarza2020}. The eDFs considered are the Maxwell-J\"uttner distribution for thermal electrons and the $\kappa$ distribution for non-thermal electrons, where the thermal distribution is given by:}

\begin{equation}
\frac{dn_{\rm e}}{d\gamma_{\rm e}} = \frac{n_{\rm e}}{4 \pi \Theta_{\rm e}} \frac{\gamma_{\rm e} \sqrt{\gamma_{\rm e}^2 - 1}}{K_2\left(1/\Theta_{\rm e}\right)} \exp \left(- \frac{\gamma_{\rm e}}{\Theta_{\rm e}}\right) 
\label{eq:mjedf}, 
\end{equation}
where $n_{\rm e}$ is the electron number density, $\gamma_{e}$ is the electron Lorentz factor and $K_{2}$ is the Bessel functions of second kind. The kappa eDF consisting of a thermal core and a non-thermal tail is given by:

\begin{equation}
\frac{dn_{\rm e}}{d\gamma_{\rm e}} = \frac{N}{4 \pi} \gamma_{\rm e} \sqrt{\gamma_{\rm e}^2 - 1} \left(1 + \frac{\gamma_{\rm e}-1}{\kappa w}\right)^{-(\kappa+1)},
\label{eq:kappaedf}
\end{equation}
where $N$ is a normalisation factor and the emission and absorption coefficients can be found in \citet{Pandya2016}. 
For the kappa eDF we follow \citet{Davelaar2019} and include in the width of the kappa eDF, $w$, a fraction, $\varepsilon$, of the magnetic energy used for the acceleration of non-thermal particles: 

\begin{equation}
   w:= \frac{ \kappa -3 }{\kappa} 
   \Theta_{\rm e} + \frac{\varepsilon}{2}\left[1+\tanh(r-r_{\rm inj})\right]\, \frac{ \kappa -3 }{6 \kappa} \frac{m_{\rm p}}{m_{\rm e}} \sigma, \label{eq:w} 
\end{equation}
where $r_{\rm inj}$ corresponds injection radius of non-thermal particles with energy contribution from the magnetic field and $m_{\rm e,p}$ are the electron and proton masses. 

{The parameter $r_{\rm inj}$ is introduced as a phenomenological scale marking the onset of non-thermal particle injection into the jet. Following previous works, we adopt $r_{\rm inj} \sim 10\,M$, corresponding approximately to the location of the stagnation surface \citep{Nakamura2018}, where the flow transitions from inflow to outflow \citep{Fromm2021,Osorio2021,Zhang2024,Zhang2026}. Beyond this region, accelerated particles are advected along the jet and can contribute significantly to the emission. In Appdenix B we explore the impact of different values for $r_{\rm inj}$ on the spectral index. If not state otherwise we use $r_{\rm inj}=10\,M$ throughout this work.} {\newline In addition, we include sub-grid prescriptions for the $\kappa$ parameter that depend on $\sigma$ and $\beta_{\rm p}$, motivated by particle-in-cell (PIC) simulations. For the reconnection heating model, we adopt the parametrization of \citet{Ball2018a}, which is based on PIC simulations of collisionless magnetic reconnection in current sheets:
\begin{equation}
    \kappa = 2.8 + 0.7\sigma^{-1/2} + 3.7\sigma^{-0.19}\tanh\!\left(23.4\,\sigma^{0.26} \beta_{\rm p}\right),
    \label{eq:kapparecon}
\end{equation}
while for the turbulent heating model we use the prescription of \citet{Meringolo2023}, derived from PIC simulations of plasma turbulence:
\begin{equation}
    \kappa = 2.8 + 0.2\sigma^{-1/2} + 1.6\sigma^{-6/10}\tanh\!\left(2.25\,\sigma^{1/3}\beta_{\rm p} \right),
    \label{eq:kappaturb}
\end{equation}
thereby ensuring a more self-consistent connection between the adopted heating mechanisms and the resulting non-thermal electron distributions \citep[see also][]{Zhang2026}. The $\kappa$ parameter is directly related to the slope $s$ of the power-law tail of the electron distribution function, $dn_{\rm e}/d\gamma_{\rm e} \propto \gamma_{\rm e}^{-s}$, via $s = \kappa - 1$. To remain within the validity range of the synchrotron emission and absorption coefficients of \citet{Pandya2016}, we restrict $3 < \kappa \leq 8$. In practice, this condition is primarily satisfied in magnetised outflow regions (i.e. the jet and its sheath), where non-thermal particles therefore contribute to the emission. In regions where this condition is not fulfilled, we adopt a purely thermal (Maxwell-J\"uttner) electron distribution. We further distinguish between the highly magnetised jet spine ($\sigma > \sigma_{\rm cut}$) and the jet sheath ($\sigma \leq \sigma_{\rm cut}$). During the GRRT calculations, the jet spine is excluded to avoid regions potentially affected by GRMHD floor prescriptions (see Sect. \ref{sec:GRMHD}). Following \citet{Fromm2021} and \citet{Osorio2021}, we adopt $\sigma_{\rm cut}=3$. The only free parameter in our GRRT setup is the fraction of magnetic energy, $\varepsilon$, that determines the width of the $\kappa$ distribution (see Eq.~\ref{eq:w}). We fix $\varepsilon=0.5$, as this value provides the best simultaneous fit to the broadband spectral energy distribution of M87 and the observed jet width \citep[see][]{Fromm2021,Osorio2021}.}

\begin{figure}[h!]
\centering
\includegraphics[width=0.49\textwidth]{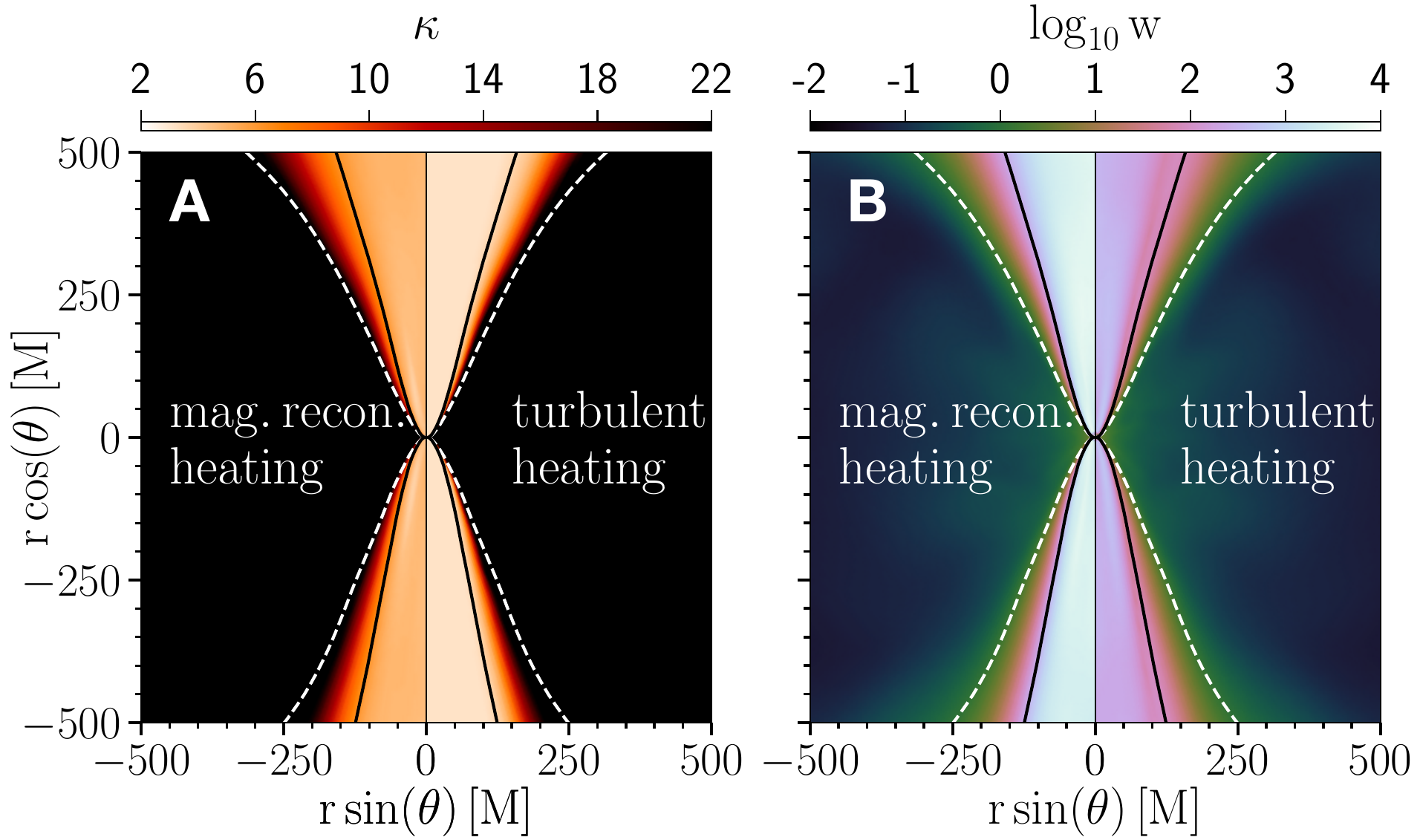}
 \caption{{Azimuthally and time-averaged ($28000\,\mathrm{M} \leq t \leq 30000\,\mathrm{M}$) distributions of the $\kappa$ (panel A) and the width of the $\kappa$-edf, $w$ (panel B). In both panels, the left half corresponds to magnetic reconnection heating, while the right half shows the turbulent heating model. Solid black lines mark $\sigma = 3$, while dashed lines indicate the boundary between bound ($-h u_t < 1.02$) and unbound ($-h u_t > 1.02$) plasma.}}
\label{fig:Morphology2}
\end{figure}

\begin{figure*}[t!]
\centering
\includegraphics[width=\textwidth]{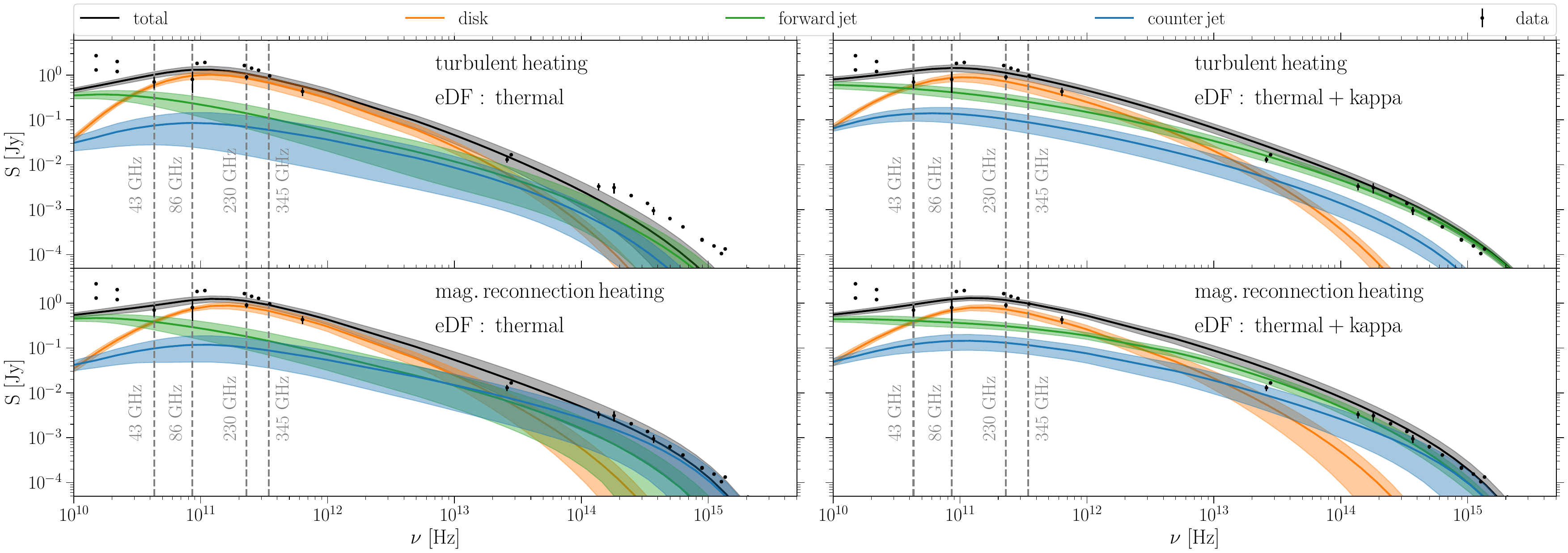}
 \caption{{ Average broad-band spectrum (using the interval $28000\,\rm{M} \leq t \leq 30000\,\rm{M}$) for turbulent heating (top) and magnetic reconnection heating (bottom) for thermal (left) and hybrid thermal-kappa eDF (right)}}
\label{fig:spectra}
\end{figure*}

{In panel A of Fig.~\ref{fig:Morphology2} we show the distribution of the $\kappa$ parameter, $\kappa$ (Eq.~\ref{eq:kapparecon} and \ref{eq:kappaturb}), and in panel B the corresponding width of the $\kappa$ distribution, $w$ (Eq.~\ref{eq:w}). In both panels, the left half corresponds to magnetic reconnection heating, while the right half shows the turbulent heating model. In the turbulent case, $\kappa$ attains smaller values in the jet spine (approaching $\kappa \sim 3$) and remains systematically lower also in the jet sheath compared to the reconnection model. This implies a flatter (i.e. shallower) non-thermal tail in these regions. Combined with the differences in the electron temperature associated with the respective heating prescriptions, this leads to noticeable variations in the width $w$ of the $\kappa$ distribution between the two models. As in the previous discussion, the highly magnetised jet spine (indicated by the black solid lines corresponding to $\sigma = 3$) is excluded in the subsequent GRRT analysis, as these regions are affected by numerical floor prescriptions. While this represents a limitation of the current approach, the distributions of $\kappa$ and $w$ outside the spine remain sufficiently distinct to capture the impact of the underlying heating models. Overall, the resulting spatial trends in both $\kappa$ and $w$ are consistent with recent studies \cite{Osorio2026}.}
{\newline For the final radiative transfer we fix  the mass, distance and viewing angle ($\vartheta=160^\circ$) of M87 while iterating the mass accretion rate to match an average flux of $1.0\,\mathrm{Jy}$ at $230\,\mathrm{GHz}$ for the compact emission region ($r \sim 100\,\mu\mathrm{as}$) over $28000\,M \leq t \leq 30000\,M$. The resulting accretion rates are $2.79\,(2.59)\times10^{-5}\,M_\odot\,\mathrm{yr^{-1}}$ for magnetic reconnection heating without/with non-thermal particles and $2.72\,(2.47)\times10^{-5}\,M_\odot\,\mathrm{yr^{-1}}$ for turbulent heating. In addition to the total jet--disk emission, we compute radiative transfer separately for the disk, forward jet, and counter-jet to quantify their individual contributions. For each component, emission is calculated individually while retaining the absorption from the full system \citep[see also][]{Davelaar2019}, ensuring that the total flux equals the sum of its components.}

\section{Results}
\label{sec:results}
\subsection{Broad-band spectra}
\label{sec:broadband}

{In Fig.~\ref{fig:spectra} we present the time-averaged\footnote{averaged over $2000\,M$ between $28000\,M$ and $30000\,M$} broad-band spectra for both heating models and electron distribution functions (eDFs). In addition to the total jet--disk emission (black curves), we show the decomposed contributions from the disk (orange), forward jet (green), and counter-jet (blue). All models produce a flat radio spectrum ($\nu \lesssim 10^{12}\,\mathrm{Hz}$). When including a $\kappa$ eDF, both heating models provide a good match to the NIR emission (right panels of Fig.~\ref{fig:spectra}). In fact, the resulting spectra are very similar, and no clear systematic difference between turbulent and magnetic reconnection heating can be identified at these frequencies. This indicates that, once non-thermal particles are included, the detailed heating mechanism becomes largely degenerate in the integrated spectral properties. Although the total spectra are broadly similar across all models, the component decomposition reveals systematic differences. For purely thermal eDFs, the disk-dominated frequency range extends from $2\times10^{10}$\,Hz to $6\times10^{13}$\,Hz for turbulent heating and from $4\times10^{10}$\,Hz to $2\times10^{13}$\,Hz for reconnection heating, indicating more efficient electron heating in the jet for the latter. When including a $\kappa$ eDF, the disk contribution at low frequencies remains largely unchanged, while the disk-dominated high-frequency range decreases to $\sim10^{12}$\,Hz for both heating models. At the same time, the jet contribution increases significantly and the spectrum becomes flatter due to the presence of non-thermal particles. This effect is slightly more pronounced for magnetic reconnection heating than for turbulent heating (see $4\times10^{10}$\,Hz to $3\times10^{12}$\,Hz).}

\begin{figure*}[h!]
\centering
\includegraphics[width=0.85\textwidth]{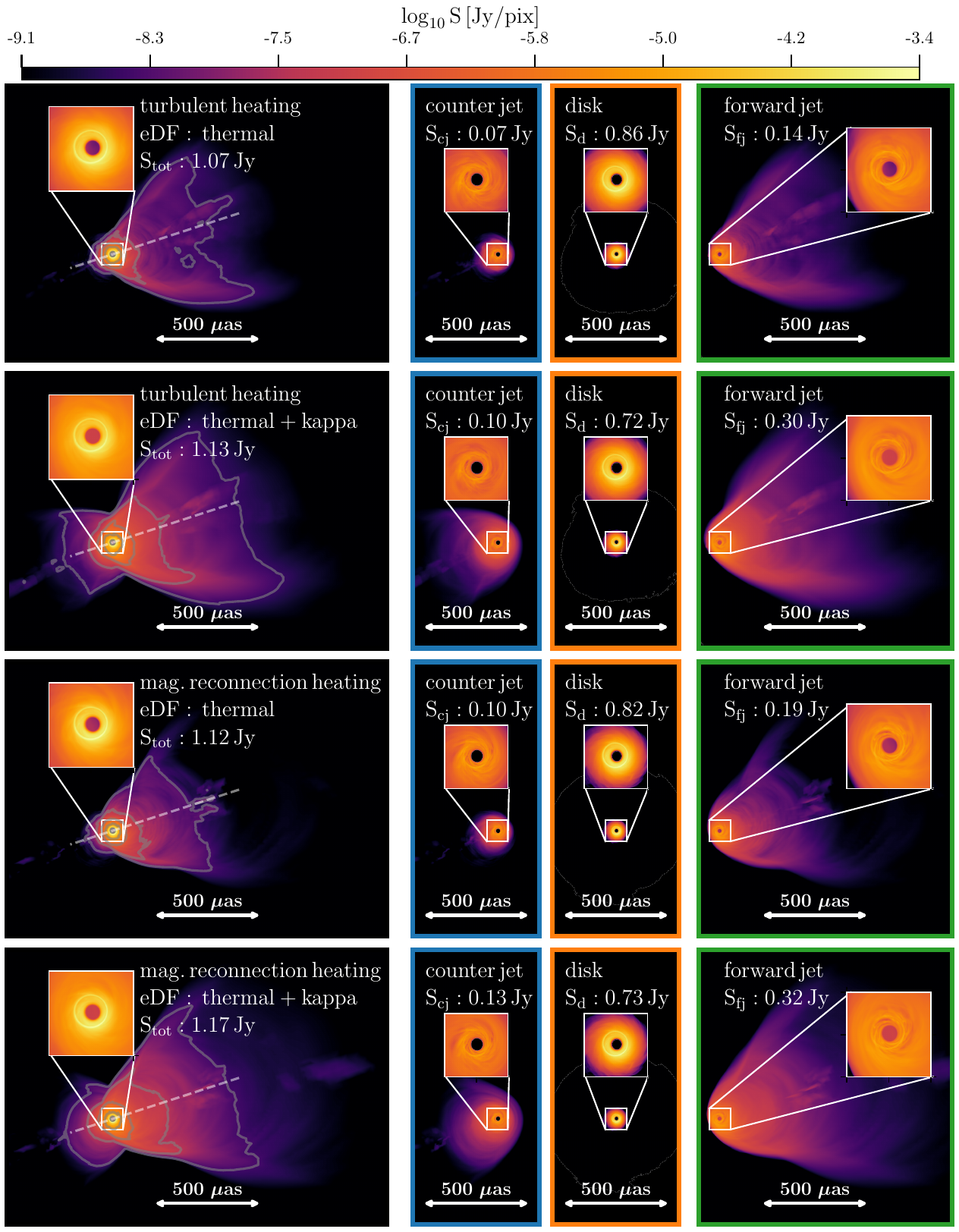}
\caption{{Average 230\,GHz images (left column) and their decomposition into counter jet, disk, and forward jet for turbulent (top two rows) and magnetic reconnection heating (bottom two rows). The first and third rows assume a thermal eDF, while the second and fourth include a thermal+ $\kappa$ eDF. Insets show zooms of the horizon-scale structure. Images are time-averaged over $28000\,M$–$30000\,M$ and computed at a viewing angle $\vartheta=160^\circ$. Grey contours (left column) denote dynamic ranges of $10^2$–$10^5$. The dashed white line in the first column corresponds to the jet axis.} }
\label{fig:230images}
\end{figure*}

\subsection{230\,GHz image structure}
\label{sec:230structure}
{In addition to the broad-band spectra, we investigate the jet--disk structure via time-averaged ($28000\,\mathrm{M} \leq t \leq 30000\,\mathrm{M}$) 230\,GHz images (Fig.~\ref{fig:230images}). The left column shows the total emission and its decomposition into counter jet, disk, and forward jet, with component fluxes summing to the total. The first two rows correspond to turbulent heating (thermal and thermal+$\kappa$ eDF), while the third and fourth rows show magnetic reconnection heating. Insets display horizon-scale structures ($\sim100\,\mu$as), and contours indicate dynamic ranges of $10^2$--$10^5$. All models produce wide jets extending to $\sim1000\,\mu$as. On horizon scales ($\sim50\,\mu$as), the emission is dominated by a bright ring with a central flux depression, primarily originating from the disk, while the forward jet contributes to the central region. At 230\,GHz, the disk dominates the total flux. The relatively small contrast between forward and counter jet emission suggests that Doppler boosting is partly mitigated by optical depth and geometric effects. Turbulent heating yields smoother, edge-brightened emission concentrated in the jet sheath, consistent with enhanced heating in low-$\beta$ regions. Reconnection heating produces more structured emission patterns, although similar arc-like features are also present in the turbulent case, and a direct association with magnetic flux tubes cannot be firmly established within the present sub-grid framework. Including a $\kappa$ eDF introduces a bright counter jet and extends the jet emission on intermediate scales ($100\,\mu\mathrm{as} \lesssim r \lesssim 400\,\mu\mathrm{as}$), while leaving the disk largely unchanged. This reflects the association of non-thermal particles with magnetised outflows, whereas the disk remains thermally dominated. The flatter spectral slope of the $\kappa$ distribution as compared to the thermal distribution enhances high-frequency emission, producing more extended intensity in the jets.}\\

\begin{figure}[h!]
\centering
\includegraphics[width=0.5\textwidth]{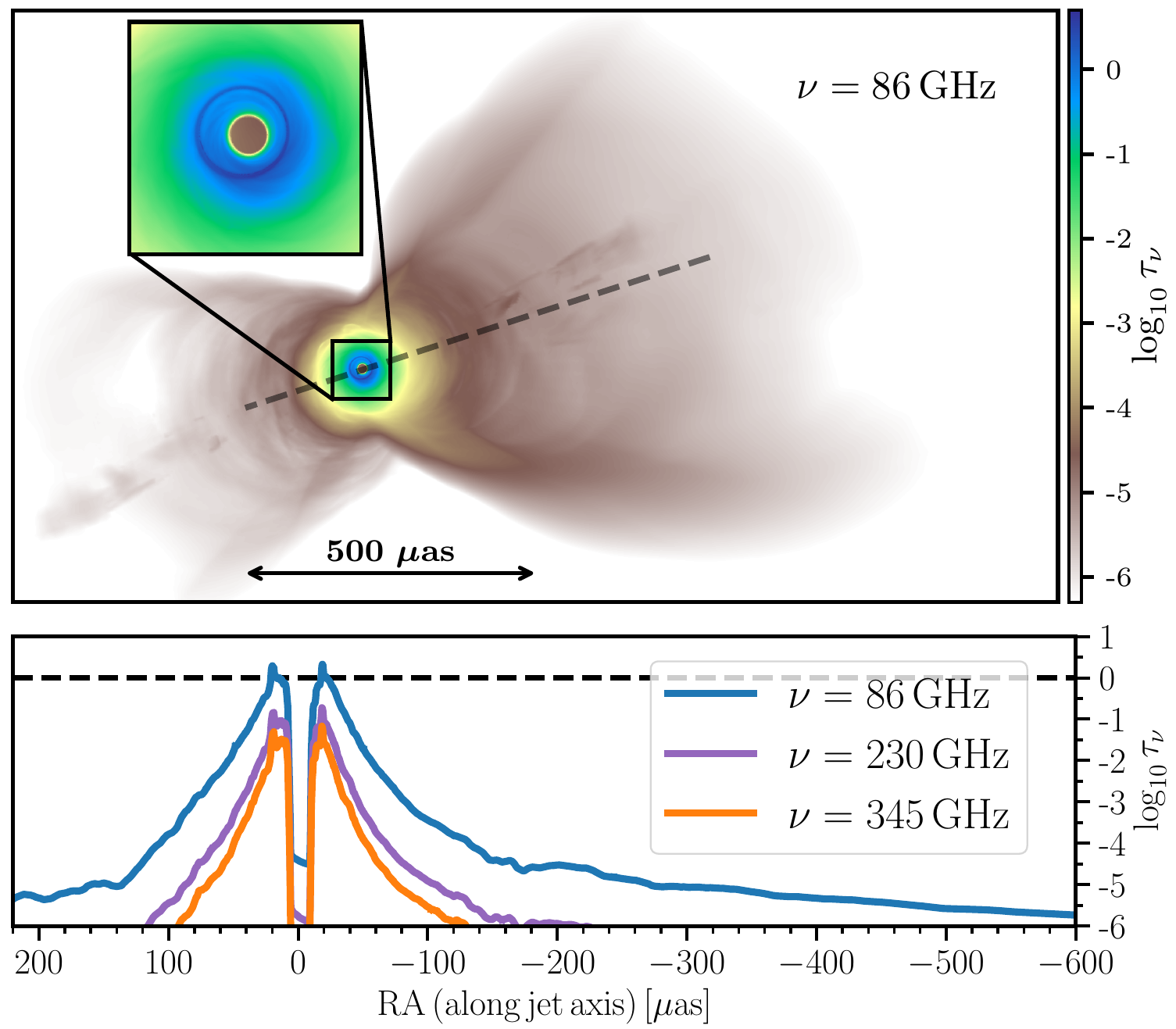}
 \caption{Time-averaged opacity distribution, $\tau_\nu$, for the turbulent heating model. Top: two-dimensional map of $\tau_\nu$ at 86\,GHz. Bottom: opacity profiles along the projected jet axis (dashed line in the top panel) for 86, 230, and 345\,GHz.}
\label{fig:opacity}
\end{figure}

\begin{figure*}[h!]
\centering
\includegraphics[width=0.85\textwidth]{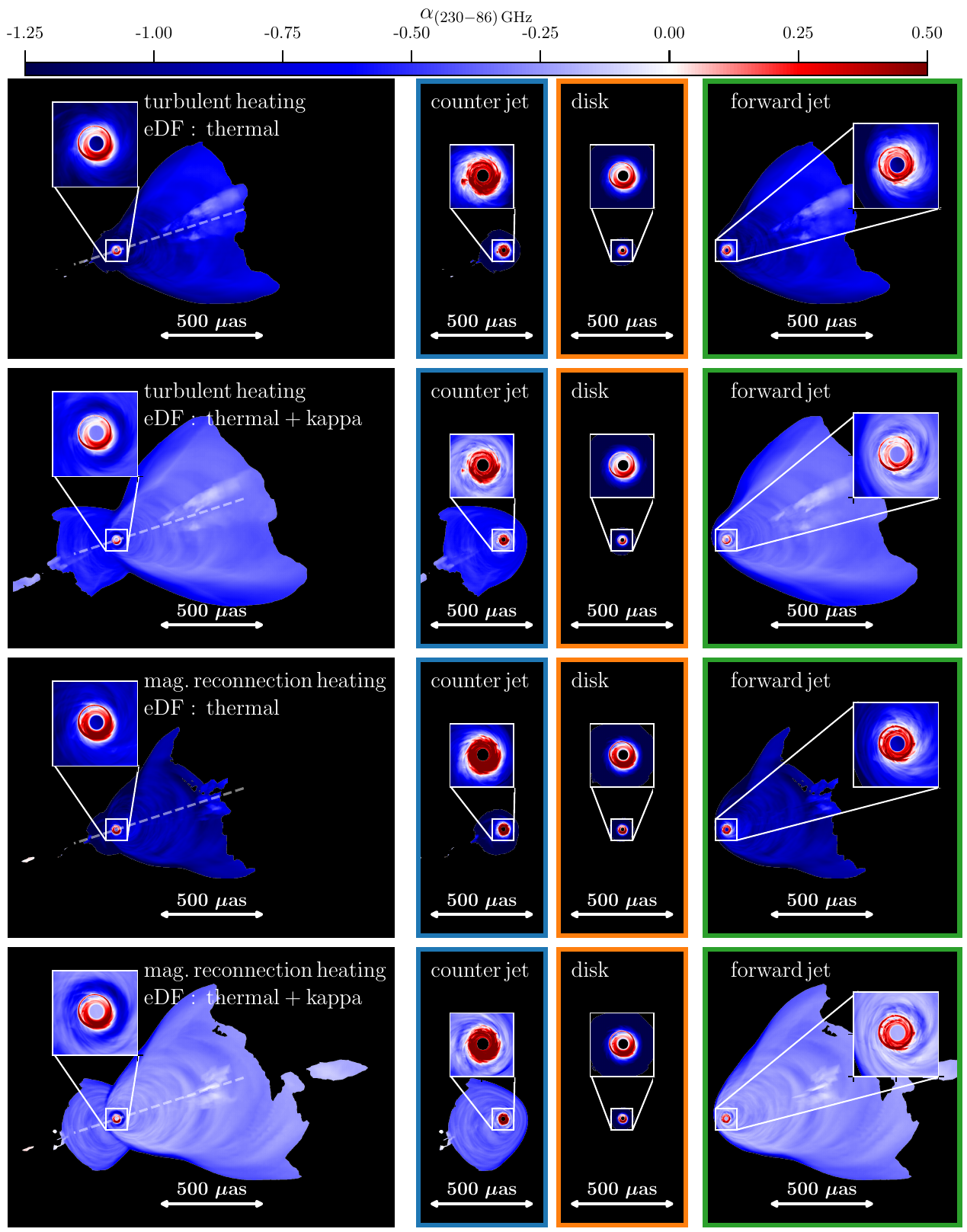}
\caption{Same as Fig.~\ref{fig:230images} but plotting averaged spectral index computed between 86\,GHz and 230\,GHz.}
\label{fig:23086spiximages}
\end{figure*}

\begin{figure*}[h]
\centering
\includegraphics[width=1\textwidth]{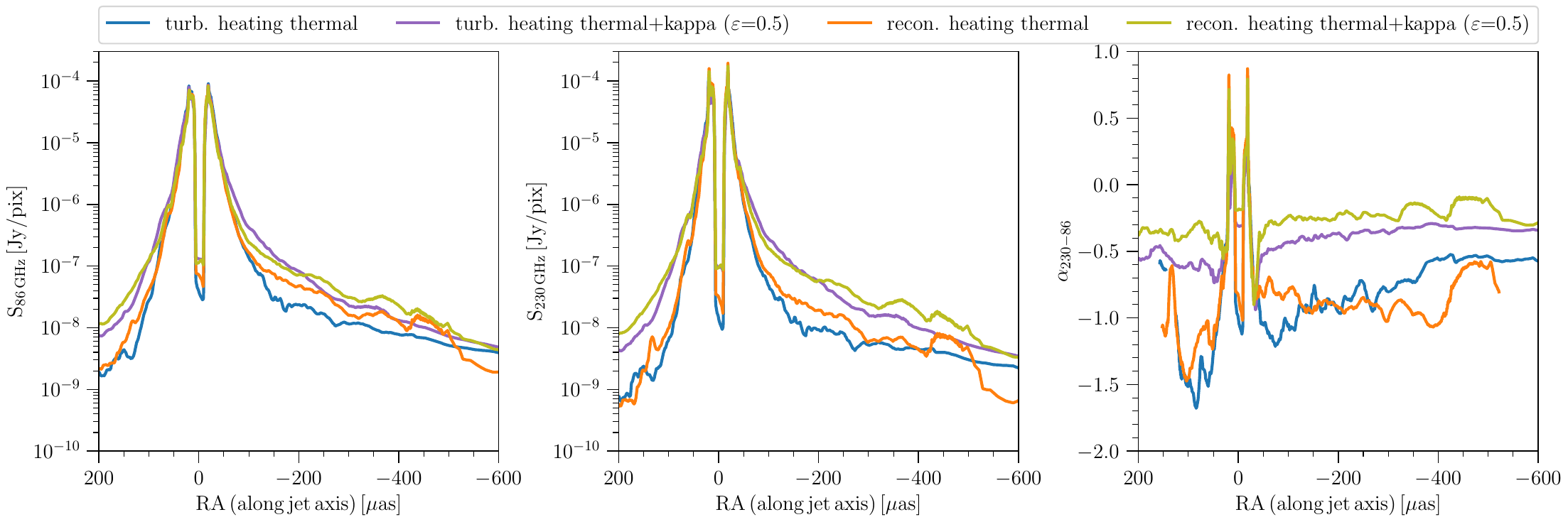}
\caption{Flux densities at 86\,GHz (left), 230\,GHz (middle) and spectral index (right) along the jet axis (dashed line in Figs.~\ref{fig:230images} and \ref{fig:23086spiximages}) for different heating mechanism and eDFs.}
\label{fig:fluxaxis}
\end{figure*}

\subsection{230\,GHz -- 86\,GHz spectral index}
{The spectral index between two frequencies, $\nu_1$ and $\nu_2$, is defined as
\begin{equation}
\alpha = {\log\left[S(\nu_1)/S(\nu_2)\right]}/{\log(\nu_1/\nu_2)},
\end{equation}
and can be used to trace the injection and acceleration of non-thermal particles as well as to distinguish between different heating mechanisms and eDFs. In Fig.~\ref{fig:23086spiximages} we present spectral index maps between 86\,GHz and 230\,GHz, decomposed into disk, forward, and counter jet components. All models exhibit an optically thick ring structure (see insets), while outside ($50\,\mu\mathrm{as} \lesssim r \lesssim 100\,\mu\mathrm{as}$) the spectral index decreases. This gradient is generally steeper for turbulent heating than for magnetic reconnection heating, independent of the eDF. \newline To assess the role of opacity, we computed in Fig. \ref{fig:opacity} optical-depth maps and profiles at 86, 230, and 345\,GHz. While the emission is largely optically thin at 230\,GHz and 345\,GHz ($\tau_\nu \ll 1$), the 86\,GHz emission reaches optical depths of order unity in the innermost disk and jet-base region. Away from the central $\sim50$--$100\,\mu\mathrm{as}$, the opacity rapidly decreases and the flow becomes optically thin. Consequently, opacity effects can influence the spectral index close to the ring and photon orbit, whereas the larger-scale spectral-index structure is primarily determined by the electron distribution function and heating prescription. \newline Within the central flux depression (photon orbit region), the spectral index is primarily determined by forward jet emission, linking horizon-scale observables to the large-scale jet. This effect requires a sufficiently large field of view to include emission in front of the black hole. On larger scales ($r \gtrsim 100\,\mu\mathrm{as}$), the impact of the eDF becomes evident: thermal models show steep spectra ($\alpha \sim -1$), while $\kappa$ models yield flatter values ($-0.25 \lesssim \alpha \lesssim 0$), consistent with the emissivities, i.e., exponential versus power-law,
\begin{equation}
j_{\nu,\rm thermal}\propto\exp\left[-\left(\frac{\nu}{B\Theta_e^2}\right)^{1/3}\right], \quad 
j_{\nu,\kappa}\propto \left(\frac{\nu}{Bw^2}\right)^{-(\kappa -2)/2}.
\end{equation}
This reflects the flatter spectral slope of the $\kappa$ distribution and enhanced high-frequency emission. All models show localised flattening of the spectral index associated with the jet spine, where higher magnetic fields and electron temperatures increase emissivity. For $\kappa$ models, this effect is further enhanced by smaller $\kappa$ values in highly magnetised regions.\\ 
While differences between heating models are subtle in total intensity, the spectral index maps reveal clearer distinctions. Magnetic reconnection heating produces systematically flatter spectra than turbulent heating on scales $50\,\mu\mathrm{as} \lesssim |r| \lesssim 150\,\mu\mathrm{as}$, independent of the eDF. This is further illustrated in Fig.~\ref{fig:fluxaxis}, which shows flux densities at 86\,GHz and 230\,GHz and the resulting spectral index along the jet axis (dashed white line in Fig.~\ref{fig:23086spiximages}). On horizon scales ($|r| \lesssim 50\,\mu\mathrm{as}$), the flux densities are very similar for all models, except within the flux depression. At larger scales ($150\,\mu\mathrm{as} \lesssim |r| \lesssim 400\,\mu\mathrm{as}$), reconnection heating yields higher fluxes, particularly for the thermal eDF, indicating more efficient electron heating, which directly translates into flatter spectral indices.In addition, arc-like features with comparatively flatter spectral indices are more pronounced in the $\kappa$ models for magnetic reconnection heating than for turbulent heating, reflecting enhanced non-thermal particle contributions in magnetised structures.}

\begin{figure*}[b!]
\centering
\includegraphics[width=\textwidth]{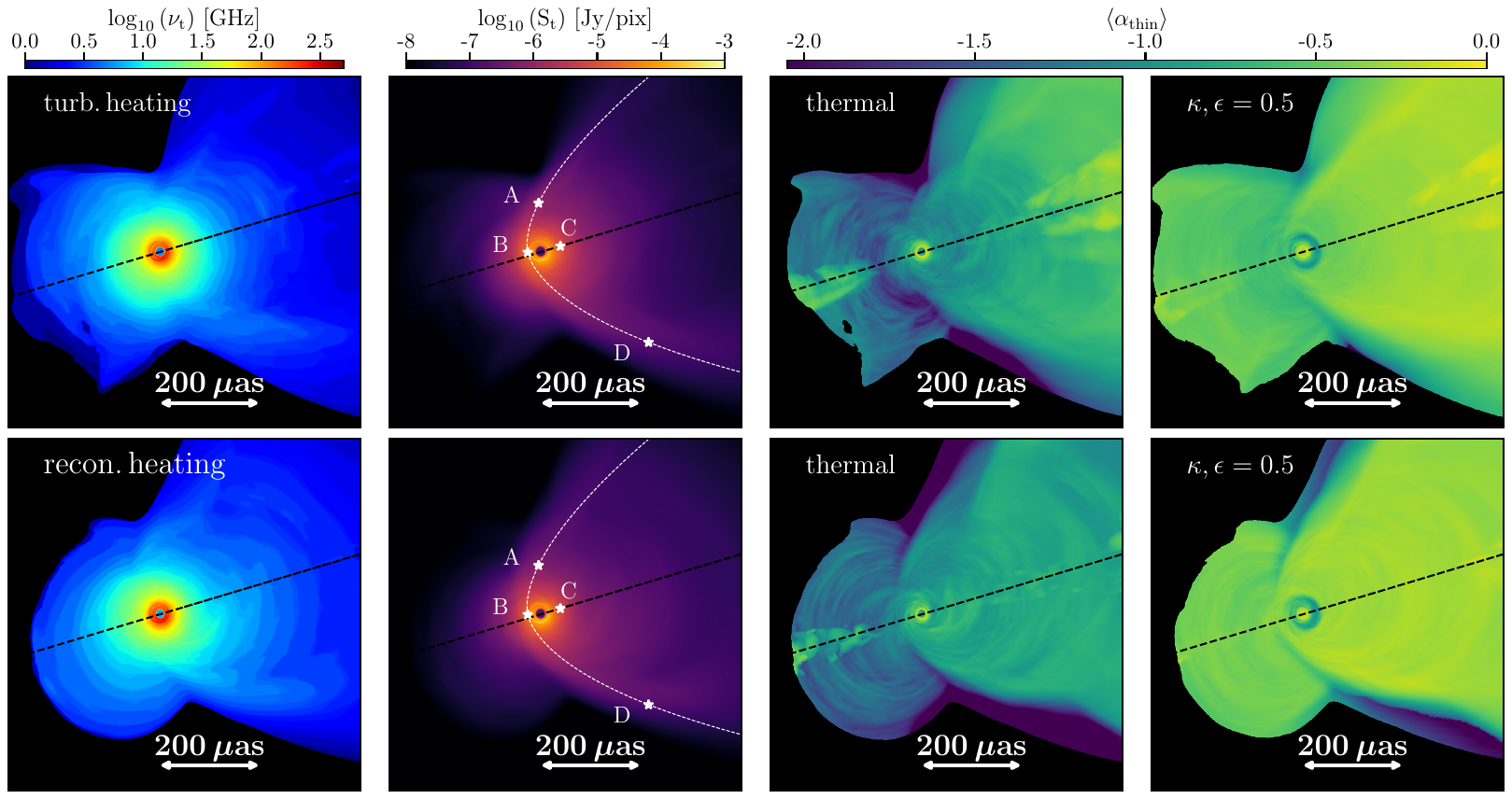}
 \caption{Distribution of the turnover frequency, $\nu_t$ (first column) and turnover flux density, $S_t$  (second column) and average optically spectral index (third and fourth column) for reconnection heating (top) and turbulent heating (bottom). The third column shows the spectral index for a thermal eDF and the fourth one for hybrid eDF consisting of thermal and kappa distribution. Notice that the turnover frequency and turnover flux density maps are averaged across eDF, namely thermal and hybrid eDF. {The dotted white curve in the second column indicates the parabolic jet profile}.}
\label{fig:turnover}
\end{figure*}

\begin{figure*}[h!]
\centering
\includegraphics[width=1\textwidth]{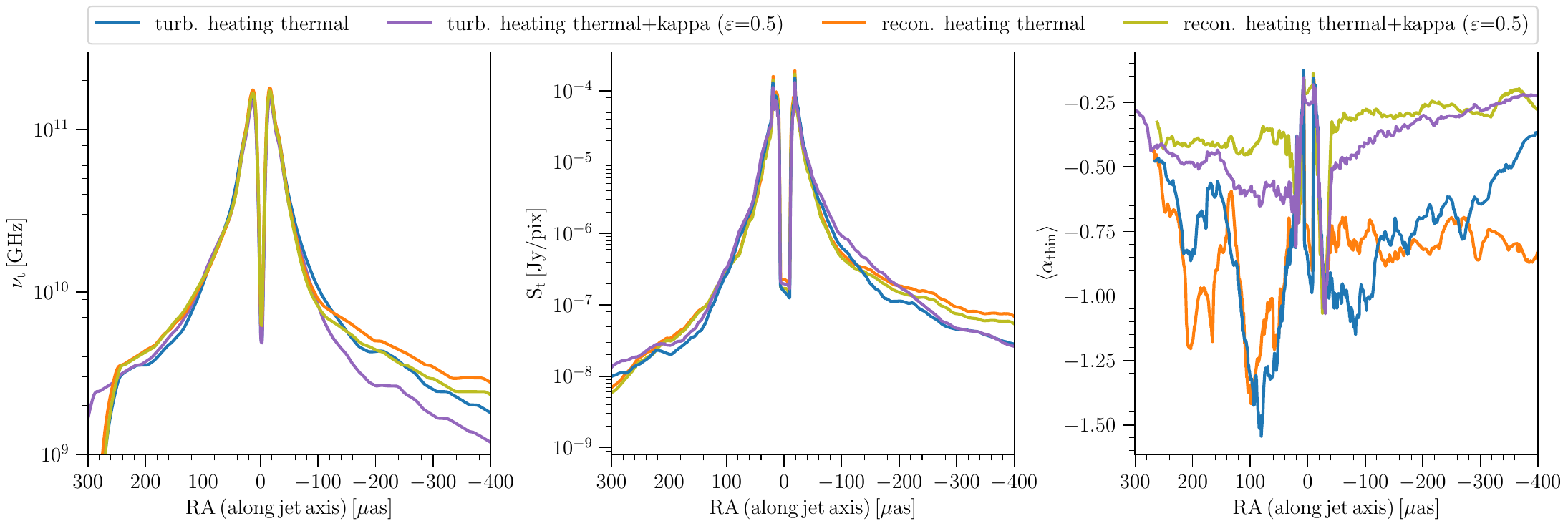}
 \caption{Turnover frequency (left), turnover flux density (middle) and average optically thin spectral index along the jet axis (dashed line in Figs.~\ref{fig:turnover})for different heating mechanism and eDFs.}
\label{fig:axialturnover}
\end{figure*}

\begin{figure}[h!]
\centering
\includegraphics[width=0.5\textwidth]{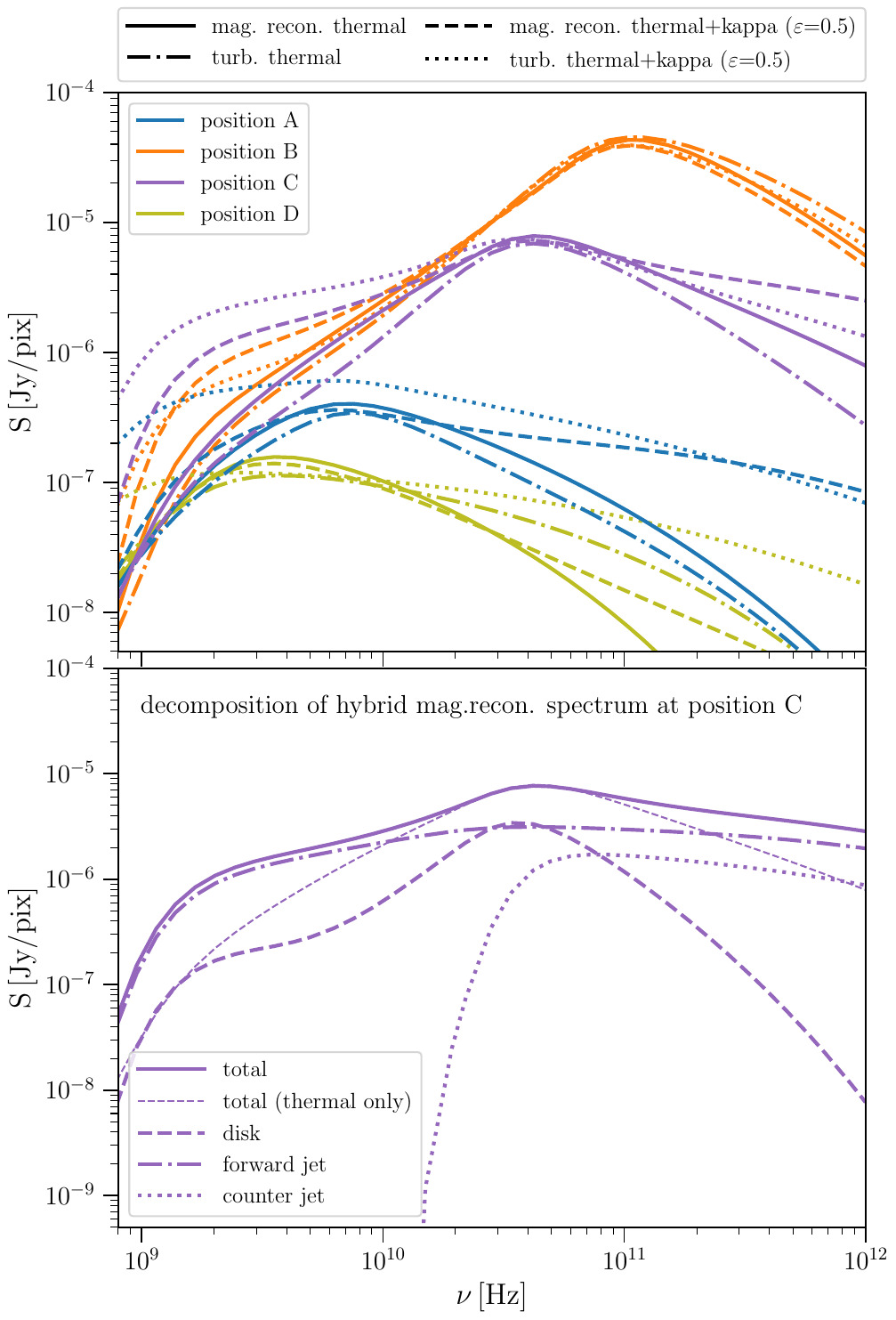}
 \caption{Individual spectra (top)  and the decomposition (bottom) for the positions indicated in Fig. \ref{fig:turnover}}
\label{fig:individualspectra}
\end{figure}

\subsection{Turnover frequency and flux density}
{Beyond the broad-band spectra and spectral-index maps, the spectral turnover provides an additional diagnostic of the plasma conditions. To characterize the local spectra, we generated GRRT images between 800\,MHz and 1\,THz using 60 logarithmically spaced frequency bins and extracted the turnover frequency, $\nu_t$, turnover flux density, $S_t$, and average optically thin spectral index, $\langle\alpha_{\rm thin}\rangle$, in each image pixel. \newline The results are presented in Figure~\ref{fig:turnover}. Independent of the heating prescription and eDF, the turnover frequency exhibits a remarkably similar morphology with peaks near the photon ring with $100\,{\rm GHz}\lesssim \nu_t \lesssim 180\,{\rm GHz}$ and decreases with distance from the black hole. The turnover flux density is concentrated toward the jet edges resampling a limb-brightened structure with parabolic shape (indicated by the dotted white line). The weak dependence of $\nu_t$ and $S_t$ on the adopted eDF reflects the fact that the spectral turnover is primarily determined by the thermal core of the particle distribution, which dominates the emissivity near the synchrotron peak. A more detailed view of $\nu_t$ and $S_t$ along the jet axis (black dashed line in Fig.~\ref{fig:turnover}) is shown in Fig.~\ref{fig:axialturnover}.The strongest differences arise in the optically thin spectral index. Thermal models yield an image-averaged values of $\langle\alpha_{\rm thin}\rangle\approx -0.75$, whereas the hybrid thermal and $\kappa$ models produce significantly flatter spectra with $\langle\alpha_{\rm thin}\rangle\approx -0.3$. This behaviour follows directly from the emissivities of the two eDFs: The thermal synchrotron emission exhibits an exponential cutoff, while the $\kappa$ distribution retains a power-law tail,
\begin{equation}
j_{\nu,\rm thermal}\propto\exp\left[-\left(\frac{\nu}{B\Theta_e^2}\right)^{1/3}\right], \quad
j_{\nu,\kappa}\propto \left(\frac{\nu}{Bw^2}\right)^{-(\kappa -2)/2}.
\end{equation}
Consequently, the $\kappa$ models maintain significantly more high-frequency emission than their thermal counterparts. In addition, the heating prescription leaves a measurable imprint on $\langle\alpha_{\rm thin}\rangle$. Magnetic reconnection heating systematically produces flatter optically thin spectra than turbulent heating within the inner $\sim100$--$150\,\mu$as, reflecting its greater heating efficiency in the magnetised inner flow. On the other hand in parts of the outer jet sheath the turbulent model yields flatter spectra, indicating more efficient particle energisation in these low-$\sigma$ regions.\newline We present representative spectra extracted at locations A--D in Fig.~\ref{fig:individualspectra} illustrating these trends. Position B is located close to the black hole and is dominated by the thermal core of the electron distribution. Therefore, all models yield nearly identical spectra, independent of both the heating prescription and the adopted eDF. The turnover frequency and flux density are essentially unchanged, demonstrating that thermal synchrotron emission dominates in this region. In addition, Position C is located at the interface between the disk and forward jet, where both components contribute significantly to the observed emission. Here the effect of the heating prescription becomes visible. Once the emission becomes optically thin, magnetic reconnection heating systematically produces higher flux densities than turbulent heating, independent of the adopted eDF, reflecting its greater heating efficiency in the inner jet-disk system. For the hybrid thermal and $\kappa$ models, this also leads to a flatter optically thin spectral slope in the reconnection case. The northern jet sheath is probed by position A located approximately $100\,\mu$as from the black hole. The thermal spectra are nearly identical for both heating models, indicating similar thermal particle populations. In contrast, the hybrid eDF models show clear differences above the turnover frequency, where turbulent heating produces both a higher flux density and a flatter optically thin spectral slope. A similar behaviour is found at position D, located in the outer southern jet sheath ($r\sim350\,\mu$as). In this low-$\sigma$ region, turbulence heats the plasma more efficiently than magnetic reconnection, resulting in enhanced high-frequency emission and flatter optically thin spectra.\newline Finally, we present a spectral decomposition at position C for the magnetic reconnection case using a hybrid eDF in the bottom panel of Fig.~\ref{fig:individualspectra}. At low frequencies ($\nu\lesssim10$\,GHz), the spectrum is dominated by the forward jet. Around the turnover frequency, the disk contribution becomes significant and shapes the spectral peak. Beyond the turnover, the thermal disk emission rapidly declines, while both the forward and counter jet contribute to the flat optically thin spectrum. The resulting high-frequency tail therefore directly traces the non-thermal particle population in the jet.}

\section{Discussion}\label{sec:discussion}
{The high-frequency emission considered in this work ($\nu \gtrsim 86$\,GHz) is largely optically thin and therefore closely traces the underlying synchrotron emissivity (see Fig. \ref{fig:opacity}). For the hybrid thermal--non-thermal electron distribution, the emissivity can be approximated as}
\begin{equation}
j_{\nu,\rm tot}\propto\exp\left(-\Theta_{\rm e}^{-2/3}\nu^{1/3}\right)+\nu^{-{(\kappa-2)}/{2}}\left[\Theta_{\rm e} + \varepsilon\sigma\right]^{\kappa-2},
\label{eq:jnukapproxTe}
\end{equation}
 
{such that the observable differences between our models are ultimately driven by the electron temperature $\Theta_{\rm e}$ and the non-thermal parameters $\kappa$ and $w$. Since $\Theta_{\rm e}$ is set by the electron-heating prescriptions, the resulting images and spectra directly reflect the different spatial distributions of turbulent and reconnection heating. \newline Our results broadly agree with previous two-temperature MAD studies. Similar to \citet{Chael2018,Chael2025}, we find that turbulent and magnetic reconnection heating produce comparable broadband spectra and horizon-scale images despite  differences in the underlying electron thermodynamics. In addition, the recent work of \citet{Monika2025} found that self-consistent turbulent-heating simulations are often observationally difficult to distinguish from simpler $R-\beta_p$ prescriptions \citet{Moscibrodzka2016} at millimetre wavelengths. Together, these results suggest that horizon-scale total-intensity observations alone provide only limited constraints on the detailed electron-heating physics. \newline Despite this apparent degeneracy, systematic differences emerge in the decomposition of the emission, in the spectral indices and the multi-frequency behaviour. Independent of the heating model, the hybrid thermal--non-thermale (kappa) eDF provides a better match to the broadband spectrum of M\,87 than the purely thermal model, consistent with our previous findings based on the $R-\beta_p$ prescription \citep{Fromm2021}. While both heating prescriptions reproduce the observed radio-to-mm spectrum reasonably well, the reconnection model provides a slightly better match to the highest-frequency data (see Fig \ref{fig:spectra}). This is expected because magnetic reconnection preferentially heats electrons in highly magnetised regions, which are concentrated near the black hole and at the jet-launching site. The decomposition of the spectra further demonstrates that the high-frequency emission originates predominantly in the jet and counter-jet, where the differences between the heating prescriptions are largest. \newline The spatial distribution of the heating becomes more apparent in the 230\,GHz images. The turbulent prescription produces enhanced emission in the low-$\beta_p$ jet sheath, leading to a more edge-brightened appearance, while the reconnection prescription preferentially heats the highly magnetised inner flow. Although arc-like structures are visible, similar patterns appear in both models and cannot be uniquely associated with magnetic flux tubes within the present sub-grid framework. Rather, these structures likely reflect underlying magnetised flow features that are selectively illuminated by the heating prescriptions. The inclusion of non-thermal particles primarily increases the relative jet contribution and extends the visible jet emission, whereas the disk emission remains largely unchanged because it is dominated by thermal electrons (see Fig. \ref{fig:230images}). \newline The strongest observational differences arise in the spectral-index maps as can be seen in Fig. \ref{fig:23086spiximages}. While all models exhibit an optically thick ring, i.e., horizon region and an optically thin jet, magnetic reconnection heating systematically produces flatter spectral indices within $\sim50 -150\,\mu$as of the black hole, best seen along the jet axis in Fig. \ref{fig:fluxaxis}. This behaviour follows directly from the higher electron temperatures reached in strongly magnetised regions. The effect is further amplified when non-thermal particles are included, since the $\kappa$ emissivity depends both on the electron temperature and on the local magnetisation through $\kappa(\beta_p,\sigma)$ and $w$. In particular, the reconnection heating models including a hybrid thermal--non-thermal eDF exhibit more extended regions of flatter spectral index and more pronounced arc-like spectral-index structures than the corresponding turbulent-heating models. \newline The turnover analysis provides a complementary view. Remarkably, the turnover frequency and turnover flux density are largely insensitive to both the heating prescription and the adopted eDF (see Fig. \ref{fig:turnover}). In all models, the turnover frequency peaks near the photon ring at $\nu_t\sim100$--$180$\,GHz and decreases gradually with distance from the black hole. This similarity reflects the fact that the turnover is controlled primarily by the thermal core of the electron distribution. In contrast, the optically thin spectral index remains strongly sensitive to both the heating model and the presence of non-thermal particles. Thermal models yield average values of $\langle\alpha_{\rm thin}\rangle\approx-0.75$, whereas hybrid models produce significantly flatter spectra with $\langle\alpha_{\rm thin}\rangle\approx-0.3$. \newline The individual spectra shown in Fig. \ref{fig:individualspectra} reveal how these differences arise. Close to the black hole (position B), all models produce nearly identical spectra because the emission is dominated by the thermal core of the electron distribution. At the disk--jet interface (position C), magnetic reconnection heating produces systematically higher optically thin fluxes and flatter spectral slopes, reflecting its greater heating efficiency in the inner jet-disk system. In contrast, in the outer jet sheath (positions A and D) turbulent heating generates higher fluxes and flatter optically thin spectra.The decomposition of the spectrum at position C shows that the flat high-frequency tail is produced jointly by the forward and counter jet, while the thermal disk component contributes primarily near the turnover frequency and rapidly declines at higher frequencies. \newline Taken together, these results indicate that total-intensity observables at 230\,GHz are relatively insensitive to the detailed electron-heating prescription and eDF, consistent with previous MAD studies. In contrast, spectral-index maps, turnover diagnostics, and resolved jet emission provide substantially greater discriminatory power. Future multi-frequency EHT, GMVA and ngEHT observations combining horizon-scale imaging with spectral-index measurements therefore offer a promising route to constraining both electron-heating and particle-acceleration mechanisms in M\,87.}

\section{Conclusion and outlook}\label{sec:conc}
We performed two-temperature 3D GRMHD simulations of black hole accretion and jet launching and analyzed the spectral properties of the resulting jet--disk system on scales up to $\sim1000\,\mu$as. Our results show that the heating prescriptions considered here, namely turbulent and magnetic reconnection heating, can be distinguished through their spectral signatures. Turbulent heating preferentially increases the electron temperature in the jet sheath, whereas magnetic reconnection heating is most efficient in the highly magnetized jet interior. These differences manifest themselves in distinct distributions and radial evolutions of the optically thin spectral index.
\newline In addition, we demonstrated that the location of non-thermal particle injection, i.e. the acceleration region, can be inferred from the spatial extent of the transition between the steep spectral indices associated with the thermal, disk-dominated emission and the flatter spectral indices associated with the non-thermal, jet-dominated emission. Future high-resolution VLBI observations will be capable of resolving these regions and may therefore provide a direct means of distinguishing between different electron-heating mechanisms through the observational signatures identified in this work.
\section{Acknowledgments}
The authors thank Oliver Porth for helpful discussions and useful comments on the manuscript.
This research is supported by the DFG research grant ``Jet physics on horizon scales and beyond" (Grant No.  443220636)
The simulations were performed on LOEWE at the CSC-Frankfurt, Iboga at ITP Frankfurt and Pi2.0 at Shanghai Jiao Tong University and on MISTRAL at the University of W\"urzburg. YM is supported by the National Key R\&D Program of China (grant no. 2023YFE0101200), the National Natural Science Foundation of China (grant no. 12273022), and the Shanghai municipality orientation program of basic research for international scientists (grant no. 22JC1410600). AN has been supported by the Hellenic Foundation for Research and Innovation (ELIDEK) under Grant No 23698. ACO is supported by DGAPA-UNAM (grant IN110522) and the Ciencia Básica y de Frontera 2023–2024 program of SECIHTI México (projects CBF2023-2024-1102 and 257435). This research is supported in part also by the ERC Advanced Grant “JETSET: Launching, propagation and emission of relativistic jets from binary mergers and across mass scales” (grant No. 884631).
 
{\it Software:} {\tt BHAC}\footnote{\href{https://bhac.science/}{https://bhac.science/}} \citep{Porth2017}, {\tt BHOSS} \citep{Younsi2020}, {\tt 
ehtim}\footnote{\href{https://achael.github.io/eht-imaging/}{https://achael.github.io/eht-imaging/}} 
\citep{Chael2018}

\bibliographystyle{aa}
\bibliography{biblio}
\clearpage
\appendix
\section{Differences in turnover values across heatings and eDFs}
\begin{figure}[h]
\centering
\includegraphics[width=0.5\textwidth]{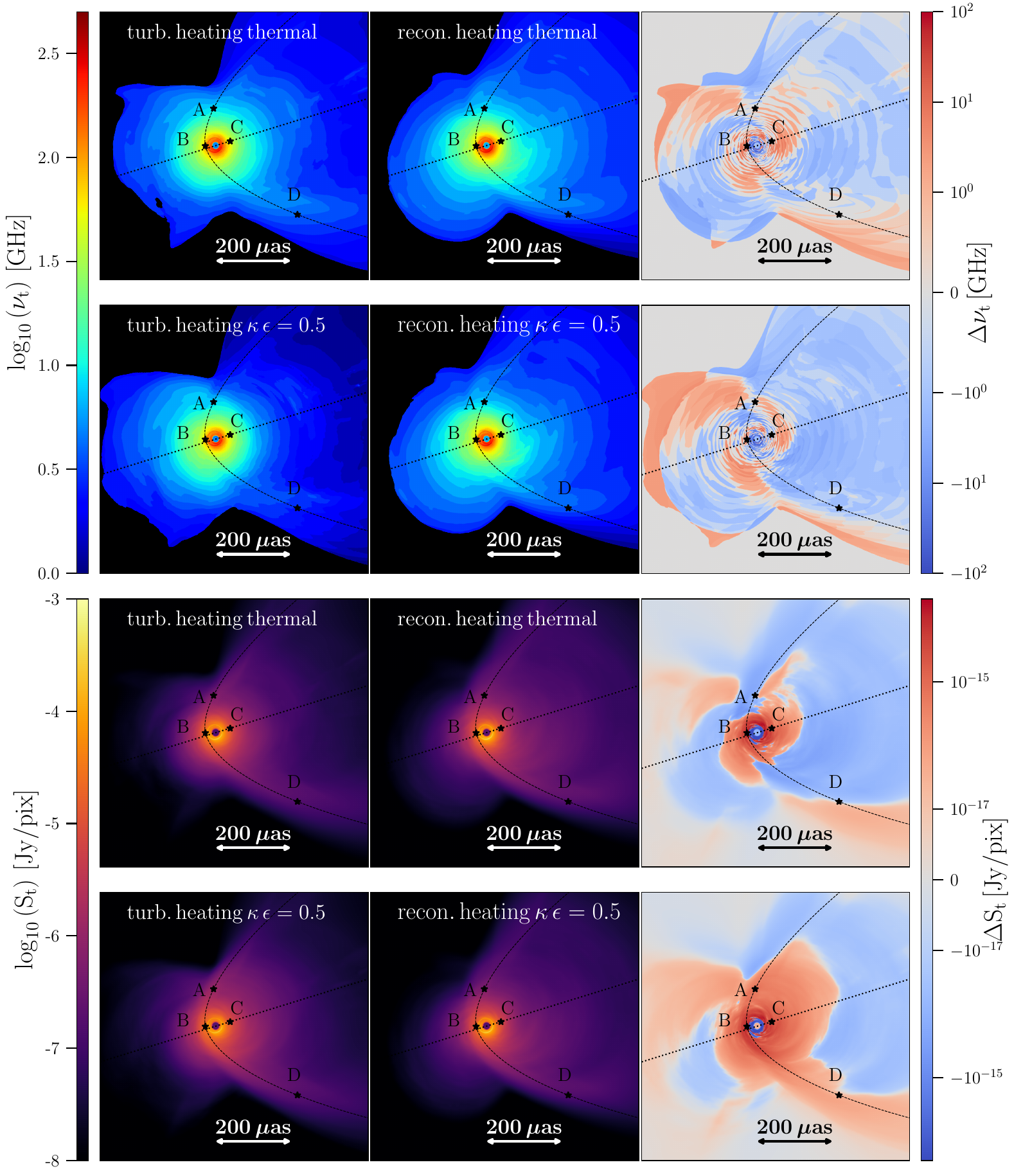}
 \caption{Distribution of the turnover frequency, $\nu_t$ (first and second row) and turnover flux density, $S_t$  (third and forth column). The difference between the first and second column is showed in the third one. The model including only a thermal eDF are shown in the first and third row and for the hybrid one in the second and forth row.}
\label{fig:diffheating}
\end{figure}

\begin{figure}[h]
\centering
\includegraphics[width=0.5\textwidth]{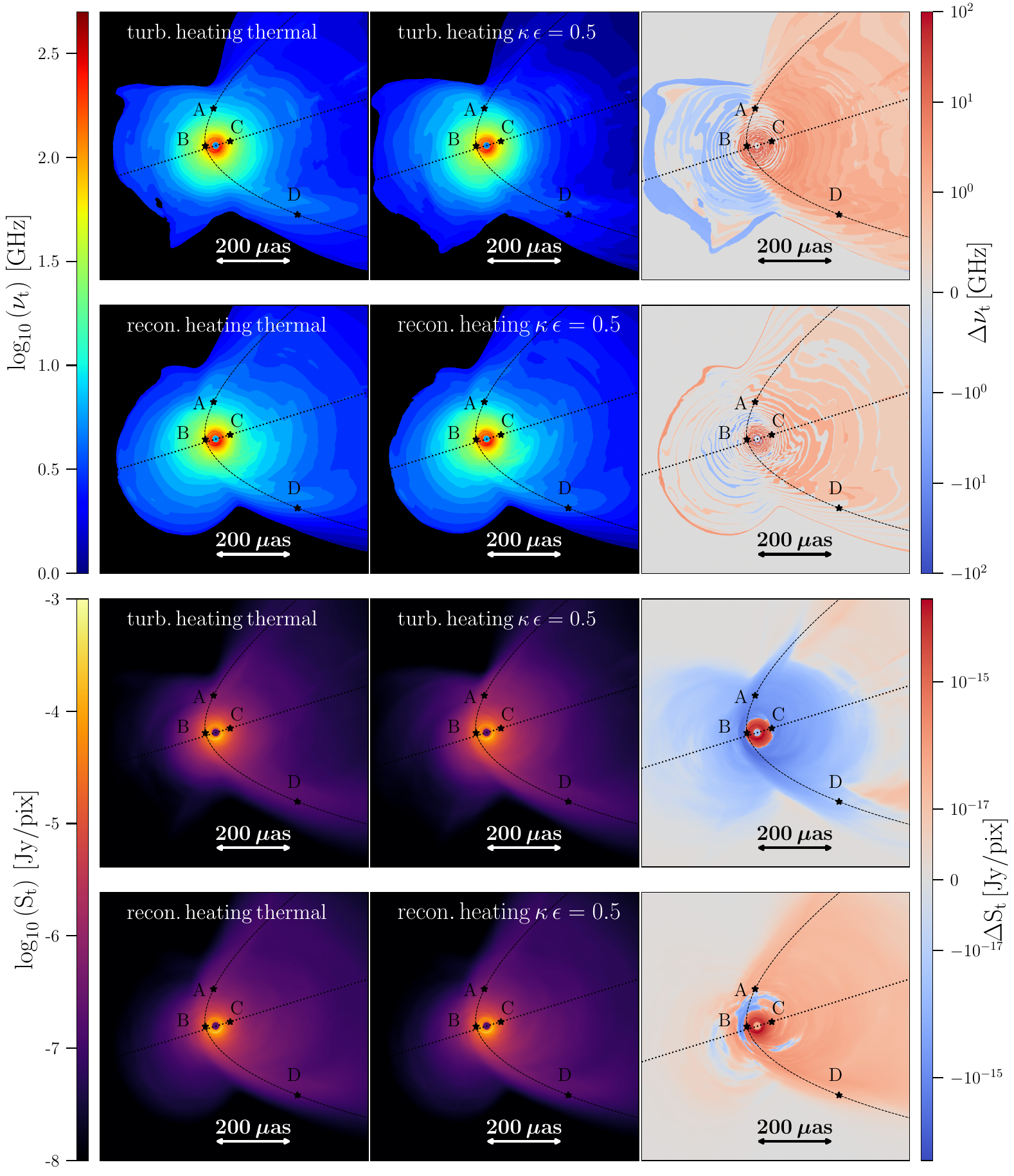}
 \caption{Distribution of the turnover frequency, $\nu_t$ (first and second row) and turnover flux density, $S_t$  (third and forth column). The difference between the first and second column is showed in the third one. The model including only a thermal eDF are shown in the first column and for the hybrid eDF in the second one.}
\label{fig:diffedf}
\end{figure}

The detailed differences across heating and eDF are presented in Fig. \ref{fig:diffheating} and \ref{fig:diffedf}. 
From the difference plots (third column in Fig. \ref{fig:diffheating}) it can be seen that the turbulent heating dominates in the southern limb of the forward jet, the northern limb of the counter jet. These region are as mentioned in Sect. \ref{sec:discussion}  predominately heated by turbulent heating which leads to a higher electron temperature and thus higher turnover frequencies (see first row in Fig. \ref{fig:diffheating}). Including a a kappa eDF in the jets is increasing the difference in these regions (see second row in Fig. \ref{fig:diffheating}). Notice that, there is only a very small difference in the turnover flux density across the different heatings and eDFs (see third and forth row in Fig. \ref{fig:diffheating}).

The difference across the eDF keeping the heating model fixed is presented in Fig. \ref{fig:diffedf}.  The differences in the turnover frequency are small within the interior of the jet except for some larger values at the jet boundary. Within the jet the differences occur mainly within the high sigma and low plasma beta regions i.e. the region where non-thermal particles are injected. The differences in the turnover flux density are negligible $\sim 10^{-15}\,\mathrm{Jy/pix}$ as can be see in the two bottom rows of Fig. \ref{fig:diffedf}. Again here the largest differences occur on scales of the injection radius, $r_{\rm inj}$ and in high sigma low plasma beta regions. 

\section{Impact of the injection radius and fraction of magnetic energy}

\begin{figure*}[h]
\centering
\includegraphics[width=\textwidth]{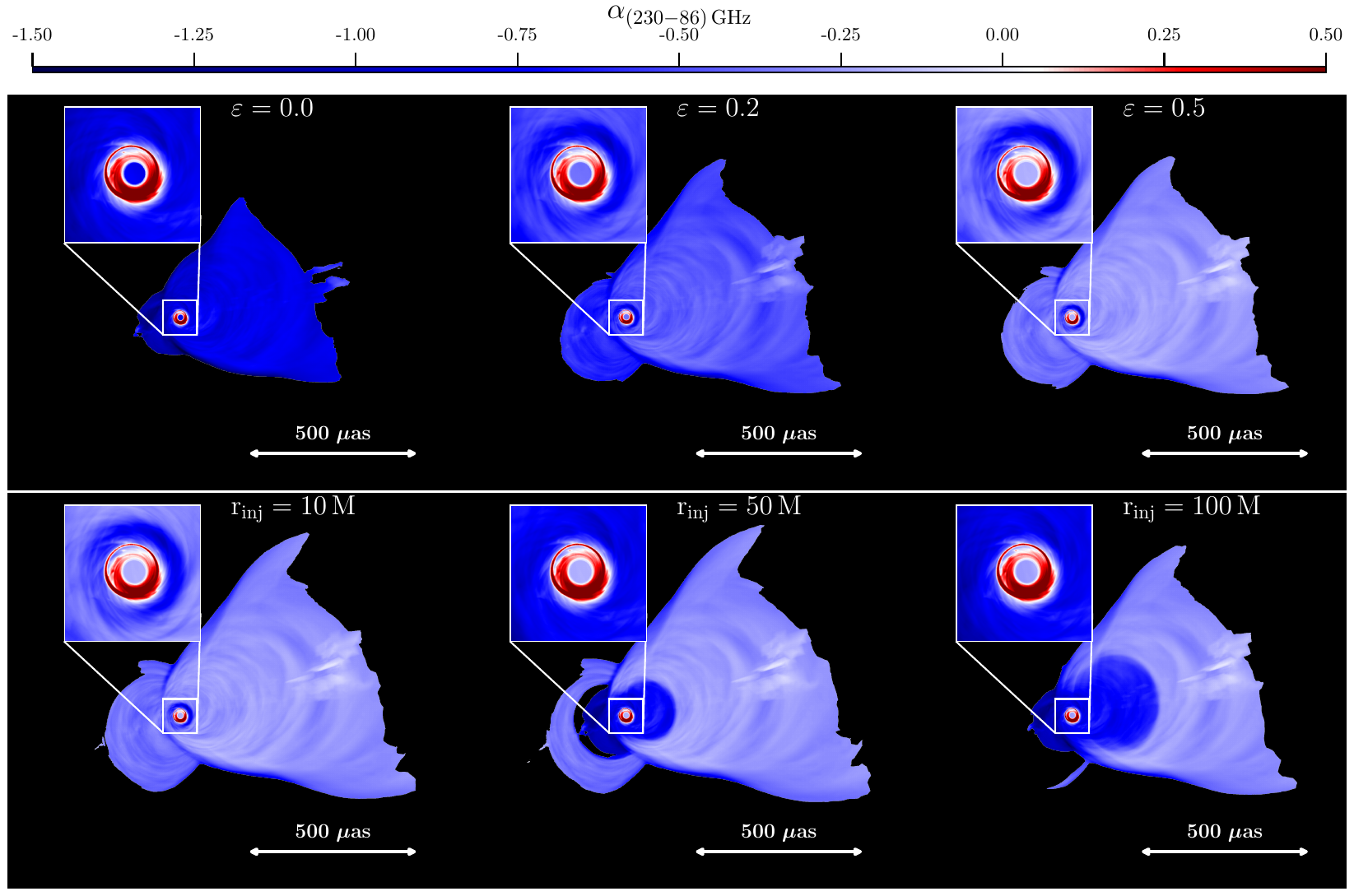}
 \caption{Spectral indices between 86\,GHz and 230\,GHz for changing fraction of magnetic field energy, $\varepsilon$, (top row) and for different values of the injection radius $\rm{r_{inj}=10,\,50,\,and\,100\,M}$ (bottom row). The images are computed for the magnetic reconnection heating model and assuming a hybrid thermal-non-thermal eDF.}
\label{fig:spixalt}
\end{figure*}

Here we investigate the impact of the injection radius, $r_{\rm inj}$, and the fraction of the magnetic energy, $\varepsilon$ used for the acceleration of non-thermal particles on the spectral indices. Since the effect across the heating models will be similar we focus on the magnetic reconnection heating model. Therefore, we performed the same GRRT calculations as in Sect.~\ref{sec:GRRT} using two additional values for the injection radius $r_{\rm inj}=50\,M$ and $r_{\rm inj}=100\,M$ while keeping the $\varepsilon=0.5$ fixed. Similarly, we used two additional values for fraction of the magnetic energy $\varepsilon=0$ and $\varepsilon=0.2$ for a fixed injection radius of $r_{\rm inj}=10\,M$. 
\newline In Figure \ref{fig:spixalt} we present the spectral indices computed between 86\,GHz and 230\,GHz. The top row displays the spectral indices computed for different values of $\varepsilon$ while keeping the injection radius fixed to 10\,M. In each panel we provide a zoom into the horizon region with a field of view of 100\,$\mu as$. The calculations show a that the spectral index in the jet region flattens with increasing $\varepsilon$. However, as mentioned previously a step spectral index region within $r_{\rm inj}=10\,M$ remains (see top row right most zoom window). 
In the bottom row we report the spectral indices for a fixed magnetic energy fraction, $\varepsilon=0.5$ while increasing the injection radius  from 10\,M (left) to 50\,M (middle) and 100\,M (right). The results show that the flat spectral index in the large scale jet $r>200\,\rm{\mu as}$ is not changed while the extent of the region with step spectral indices around the central black hole is increasing with increasing injection radius (see also the horizon scale zoom windows).

\end{document}